\documentclass[]{pasj01}
\Received{}%{yyyy/mm/dd}
\Accepted{}%{yyyy/mm/dd}
\usepackage{multirow}
\begin{document} 

\title{Detection of gamma rays around SNR HB9 and its implication to the diffusive shock-acceleration history}
%\title{Test}

%%% begin:list of authors
% Do NOT capitalize all letters in "textsc".
\author{Tomohiko Oka\altaffilmark{1}}
\altaffiltext{1}{Department of Physics, Kyoto University, Kitashirakawa Oiwake-cho, Sakyo, Kyoto 606-8502, Japan}
\email{oka.tomohiko.25n@st.kyoto-u.ac.jp}
% orcid: 0000-0002-9924-9978

\author{Wataru Ishizaki\altaffilmark{2}}
\altaffiltext{2}{Center for Gravitational Physics, Yukawa Institute for Theoretical Physics, Kyoto University, Kitashirakawa Oiwake-cho, Sakyo, Kyoto 606-8502, Japan}
\email{wataru.ishizaki@yukawa.kyoto-u.ac.jp}
% orcid: 0000-0002-7005-7139
%%% end:list of authors

%% `\KeyWords{}' always has to be placed before ``\maketitle'' 
%%  List of Key Words:  https://academic.oup.com/pasj/pages/Pasj_Keywords 

\maketitle

\begin{abstract}
%Please read ``IMPORTANT NOTICE'' carefully before preparing a manuscript. 
We analyze the GeV gamma-ray emission data  from the vicinity of the supernova remnant (SNR) HB9 (G160.9+2.6)  from the \textit{Fermi}-LAT 12-year  observations to quantify the  evolution of diffusive shock acceleration (DSA) in the SNR. 
In the vicinity of HB9, there are molecular clouds whose locations do not coincide with the SNR shell in the line of sight.
We detect significant gamma-ray emissions above 1~GeV spatially coinciding with the two prominent cloud regions, as well as a emission from the SNR shell, the latter of which is consistent with the results of previous studies.
The energy spectrum above 1~GeV in each region is fitted  with a simple power-law function of $dN/dE \propto E^{-\Gamma}$. The fitting result indicates harder spectra  with power-law indices of $\Gamma$ = $1.84 \pm 0.18$ and $1.84 \pm 0.14$ than that at the SNR shell  with $\Gamma$= $2.55 \pm 0.10$.
The  observed spectra from the cloud regions are found to be consistent with the theoretically expected gamma-ray emissions originating in the protons that  escaped from SNR HB9, where particles can be accelerated up to higher energies than those at the shell at present.
The resultant diffusion coefficient in the vicinity of the SNR is comparable to that of the Galactic mean.
\end{abstract}

%\linenumbers

\section{Introduction} \label{sec:intro}

The supernova remnant (SNR) is the most probable class of the cosmic ray (CR) acceleration site with energies up to $\sim$ PeV in our Galaxy~(see, e.g., \cite{2013A&ARv..21...70B} for a review).
However, there is no conclusive observational evidence of SNRs accelerating particles, specifically protons, up to PeV.
The lack of observational evidence of a PeV particle accelerator in SNRs against the expectation has been attempted to be explained with two scenarios: (i) an SNR is capable of accelerating particles up to PeV only  while it is young up to an age of $10^{3}$~yr~\citep{Ptuskin:2003zv},
(ii) such high-energy particles escape from the SNR at an early stage~\citep{Gabici:2006pg}.
%In order to  scrutinize the scenario that an SNR has accelerated PeV CRs in the past, it is \oldwirep{important}{essential} to quantify the  evolution of  diffusive shock acceleration (DSA) in the SNR.
In order to  scrutinize the scenario that an SNR has accelerated PeV CRs in the past, it is essential to quantify the  evolution of  diffusive shock acceleration (DSA) in the SNR.
Systematic studies exploring the population of SNRs are expected to provide information on the evolution of the CR spectra of SNRs as a function of the SNR age~\citep{2011MNRAS.410.1577O, Schure:2013kya, Gaggero:2017abc}.
However, quantifying the evolution of DSA with this kind of method is challenging due to the large diversity of the surrounding environment of SNRs~\citep{Yasuda:2019lkl, 2020PASJ...72...72S}.
\newline\indent Another way to study the  evolution of DSA is  a simultaneous observation of a single SNR, specifically its shell part  and  nearby massive molecular clouds (MCs).
If a massive cloud exists in the vicinity of the SNR, protons that escaped the SNR illuminate the cloud and  generate gamma-ray emission via the $\pi^{0}$-decay process~\citep{Aharonian:1996aa, Gabici:2006pg}.
The delay of the timing of the gamma-ray emission from the cloud region from that of the incident proton escape depends on the propagation time and accordingly reflects the particle distribution in the SNR at a specific epoch in the past.
Hence, a comparison of the spectra at the SNR shell and nearby clouds observed at roughly the same time enables us to quantify the evolution of the DSA in the SNR.
\newline\indent In observations of  SNR W28, such ``delayed'' gamma-rays have been detected from three cloud regions  with HESS and \textit{Fermi}-LAT~\citep{Aharonian:2008fp, Hanabata:2014usa}.
The observed gamma-ray spectra from these cloud regions, however, do not differ significantly from one another, and hence the maximum energy with the DSA at W28 has never been successfully measured as a function of the SNR age.
Since the angular distances from the SNR center to these individual clouds are almost the same in W28, the emissions from the clouds should originate in protons accelerated at a similar epoch.
The gamma-ray emission from another cloud (named HESS J1801$-$233) located to the north of W28 was suggested to originate in protons that have been re-accelerated by a shock-cloud interaction~\citep{2018ApJ...860...69C}, in which case the gamma-ray spectrum does not reflect the SNR age.
\citet{Hanabata:2014usa} reported detection of the fifth gamma-ray emitting region, named source W, located at the western part of W28.
Although there is a possibility that the emission is explained with protons escaping from the SNR, no cloud counterpart has been found.
Thus far, it is impossible to quantify the particle distribution at multiple epochs  with the available observations of W28.

One concern with the study using delayed gamma-ray spectra is a large dependence on the diffusion coefficient of CRs.
The diffusion coefficient may be modified in the vicinity of an SNR due to the effect of self-confinement and/or magnetic-field amplification caused by the generation of turbulent plasma waves~\citep{Wentzel:1974cp, Fujita:2011wq, DAngelo:2017rou}.
Although delayed gamma-ray emissions are detected in the vicinity of several SNRs~(e.g., \citet{Uchiyama2012ApJ}), the derived diffusion coefficients had large uncertainty, and so did the estimated amount of particle acceleration at the SNR shell in the past.
This is mainly because the SNRs discussed so far are relatively old.
For older SNRs, MCs and SNR shells often interact directly, and as a result, the observed data do not provide much information on the diffusion coefficient.
Furthermore, even if MCs are significantly separated from SNR shells, the delayed gamma-ray spectra do not differ significantly from those of the current SNR shells, and thus the data are not of much help for restricting the diffusion coefficients~\citep{Uchiyama2012ApJ}.

% \newline\indent 
Here, we focus on SNR HB9 (G160.9+2.6), which is relatively young ($\sim6.6\times10^{3}$~yr;~\citet{Leahy:2006vs}) compared  with other objects where delayed gamma-rays have been  observed.
HB9 has two additional advantages for this type of study in the DSA evolution.
Firstly, there are MCs in the vicinity of this SNR, but their locations do not coincide with the SNR in the line of sight~\citep{Sezer:2019fue}, enabling us to simply use the distance between the SNR and clouds to calculate the diffusion time.
Secondly, HB9 has observable gamma-ray emission from the SNR shell~\citep{Araya:2014kra}, which is essential to estimate the current maximum energy of the accelerated particles at the SNR shock.

HB9 has a large angular size (2$^{\circ}$ in diameter) and a shell-type radio morphology (e.g., \citet{Leahy:1991ajl}) with a radio spectral index of $\alpha = -0.47 \pm 0.06$ in frequencies between $408$ and $1420$ MHz~\citep{Leahy:2006vs}.
\citet{Leahy:1995aa} estimated  its distance and age to be $1.5$~kpc and  ($8$--$20$)$\times10^{3}$~yr, respectively, using X-ray data observed with ROSAT.
Its kinematic distance was estimated to be $0.8 \pm 0.4$~kpc from H$_{\mathrm{I}}$ observations, yielding the Sedov age estimate of $6.6\times10^{3}$~yr~\citep{Leahy:2006vs}.
\citet{Sezer:2019fue} found an H$_{\mathrm{I}}$ shell expanding toward the SNR  with a velocity ranging between  $-10.5$ and $+1.8$~km s$^{-1}$ and derived a kinematic distance to be $0.6 \pm 0.3$~kpc, which is consistent with the estimate by \citet{Leahy:2006vs}.
In the gamma-ray band, spatially extended emission was detected from SNR HB9 along with its radio shell with the \textit{Fermi}-LAT $5.5$-year observations in the energy band above $0.2$~GeV~\citep{Araya:2014kra}.
The gamma-ray spectrum  was characterized by a power law  with a photon index of $1.44 \pm 0.25$ accompanied by an exponential cutoff at  $1.6 \pm 0.6$~GeV.
The spectrum is  explained with emission from inverse Compton (IC) scattering of electrons with a simple power-law  energy spectrum with a differential index of 2 and maximum electron energy of 500~GeV.
Furthermore, \citet{Sezer:2019fue} analyzed the \textit{Fermi}-LAT 10-year data in  an energy range between $0.2$  and $300$ GeV and newly detected a point-like source near the SNR shell, named PS J0506.5+4546.
The spectrum of the point source can be characterized by a simple power-law function with an index of $1.90 \pm 0.19$.  The flux  was $(6.59 \pm 3.47) \times 10^{-10}~{\rm cm^{-2} s^{-1}}$.
PS J0506.5+4546 was, however, suggested not to be related to the SNR shell, and its origin is unclear~\citep{Sezer:2019fue}.

In this paper, we study the gamma-ray morphology of SNR HB9 and the spectra of the SNR shell and the nearby cloud regions, using 12-year observations with the \textit{Fermi}-LAT, to quantify the  evolution of DSA.
In addition, since we suspect that the gamma-ray emission from PS J0506.5+4546 originates in a MC located outside the radio shell, we search for spatial correlation between the $^{12}$CO ($J=1-0$) line and gamma-ray emissions.
In Section~\ref{section:FermiObs}, we describe the \textit{Fermi}-LAT observations and show the results of our analysis. 
We present the modeling study to interpret the origin of the observed gamma-ray in Section \ref{section:Discussion}.
Summary and conclusions are given in Section \ref{section:Summary}.

\section{Gamma-ray observations} \label{section:FermiObs}
\subsection{Data selection and analysis} 

The \textit{Fermi}-LAT is capable of detecting gamma-rays in an energy range from $\sim$ 30~MeV to $>$ 300~GeV~\citep{Atwood:2009ez}.
%We analyze its 12-year data from 2008 August to 2020 August in the vicinity of SNR HB9 for an energy range of 1--500~GeV.
We analyze its 12-year data from 2008 August to 2020 August in the vicinity of SNR HB9.
The standard analysis software, Science Tools (version v11r5p3\footnote{https://fermi.gsfc.nasa.gov/ssc/data/analysis/software}), is used.
The `Source' selection criteria and instrument responses (\verb|P8R2_SOURCE_V6|\footnote{https://fermi.gsfc.nasa.gov/ssc/data/Cicerone/}) are chosen, considering a balance between the precision and photon-count statistics.
The zenith-angle threshold is set to 90~degrees to suppress the contamination of the background from the Earth rim.
We employ the tool \verb|gtlike| (in the binned mode), using a standard maximum likelihood method~\citep{Mattox:1996zz}, for spatial and spectral analyses.
% p5 l10 : "center coinciding with taht of HB9"   Please add the coordinates of the center.   
We choose a square region of 15$^{\circ}$ $\times$15$^{\circ}$ aligned with the Galactic coordinate grid with the center coinciding with that of HB9 (Ra=75.25$^{\circ}$, Dec=46.73$^{\circ}$) as the region of interest (ROI) for the (binned) maximum likelihood analysis based on Poisson statistics.
The pixel size is 0.1$^{\circ}$.
\newline\indent The source spatial-distribution model includes all the sources in the fourth Fermi catalog (4FGL; \citet{Fermi-LAT:2019yla}) within the ROI and the two diffuse backgrounds, the Galactic (\verb|gll_iem_v7.fits|) and extragalactic  (\verb|iso_P8R3_SOURCE_V2_v1.txt|) diffuse emissions. 
Regarding the emission of the SNR shell, \citet{Araya:2014kra} and \citet{Sezer:2019fue} concluded that the radio template produced with the 4850~MHz radio continuum data from the Green Bank Telescope~\citep{Condon1994AJ} is the best spatial model.
Accordingly, we use the radio template provided in the Science Tools as the spatial model.
In the fitting of the maximum likelihood analysis, all spectral parameters of HB9 SNR itself, 4FGL sources~\citep{Fermi-LAT:2019yla} located within 5$^{\circ}$ from the center of HB9, and the two diffuse backgrounds are allowed to vary freely.
We do not use the data below 1 GeV in this analysis since the fitting results of delayed gamma-ray emission in this band suffer from the systematic uncertainty in the Galactic diffuse background model~\citep{2012ApJ...750....3A}.
Note that the flux of the Galactic diffuse emission at the MC region is larger in this energy range than that of the delayed gamma-ray emission from HB9, which will be estimated in Section \ref{section:Discussion}.
\newline\indent The significance of a source is represented in this analysis by the Test Statistic (TS) defined as $-2 \mathrm{log}(L_{0}/L)$, where $L_{0}$ and $L$ are the maximum likelihood values for the null hypothesis and a model including additional sources, respectively~\citep{Mattox:1996zz}.
The detection significance of the source can be approximated as $\sqrt{\mathrm{TS}}$ when the number of counts is sufficiently large.

\subsection{Morphology study}
\label{section:FermiAna}

\begin{figure*}
    \begin{center}
    \includegraphics[width=17cm]{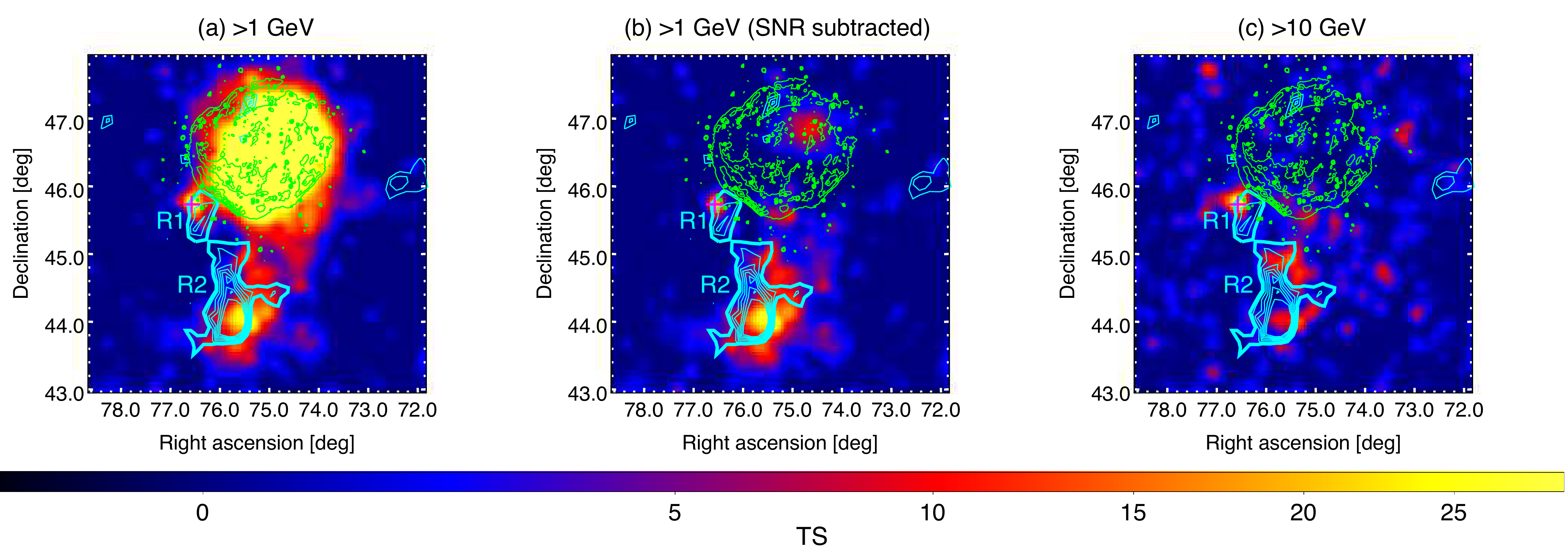}
    \end{center}
    \caption{
    Gamma-ray TS maps in the vicinity of SNR HB9 observed with the \textit{Fermi}-LAT.
    All maps are given with square bins of 0.05$^{\circ}$, and  Gaussian smoothing with a kernel $\sigma$ = 0.1$^{\circ}$ are applied.
    The energy ranges are (Panels\ a and b) 1 to 500~GeV and (c) 10 to 500~GeV.
    The subtracted background consists of (a and c) the Galactic and extragalactic diffuse emissions and known gamma-ray source contribution and (b) additionally the estimated gamma-ray emission from the HB9 SNR shell (see text for detail).
    Green contours show the radio emission of SNR HB9 at 1420 MHz with DRAO \citep{Taylor:2003ajt} and are linearly spaced in increments of 0.5 K from 5.5 K to 10.0 K.
    Cyan contours show the $^{12}$CO (J=1-0) line intensity integrated over a velocity range  between $-$10.4 and +2.6 km~$\textrm{s}^{-1}$ and are linearly spaced in increments of 1.0 K~km~$\textrm{s}^{-1}$ from 4.5 to 10.5 K~km~$\textrm{s}^{-1}$.
    The two apparent CO-emission regions (R1 and R2) are indicated.
    The magenta cross in each panel indicates the position of PS J0506.5+4546~\citep{Sezer:2019fue}.
    }
    \label{fig:Fig1_TSmap}
\end{figure*}

Figures~\ref{fig:Fig1_TSmap}(a) and \ref{fig:Fig1_TSmap}(c) show the background-subtracted gamma-ray TS map created from the \textit{Fermi}-LAT 12-year data above 1~GeV, where the background model consists of the Galactic and isotropic extragalactic emissions and the contributions from the known Fermi sources (see the previous subsection). 
The map is overlaid with cyan contours of the $^{12}$CO ($J=1-0$) line emission from the Dame survey data~\citep{Dame:2000sp}, which are integrated over a velocity range between $-10.4$ and $+2.6$~km s$^{-1}$, and also green contours of 1420 MHz radio continuum emission obtained from the CGPS survey with DRAO \citep{Taylor:2003ajt}.
In Figure~\ref{fig:Fig1_TSmap}(c), no significant emission from the SNR shell was found (green contours in the figure), which is consistent with the fact that the SNR spectrum has a cutoff below 10~GeV~\citep{Araya:2014kra}.
Figure~\ref{fig:Fig1_TSmap}(b) is a similar gamma-ray TS map to Figure~\ref{fig:Fig1_TSmap}(a), but with the contribution from the SNR shell is additionally subtracted, for which we use the 4850~MHz radio-template model as a spatial model and assume a simple power-law spectrum.
The gamma-ray excess in Figure~\ref{fig:Fig1_TSmap}(b, c) appears to be more extended than the point source J0506.5+4546 reported in the previous study and rather spatially coincident with the CO line emission.

We test whether the gamma-ray ($>$ 1~GeV) spatial distribution is correlated with CO line emissions using the likelihood method.
The $^{12}$CO ($J=1-0$) line image exhibits two distinctive regions (thus two MCs), designated as R1 and R2 (Figure~\ref{fig:Fig1_TSmap}).
For the test, we create a CO template for each of R1 and R2, which is made from the $^{12}$ CO ($J=1-0$) line image~\citep{Dame:2000sp} integrated over a velocity range between $-$10.4 and +2.6~km~$\textrm{s}^{-1}$ and cut with a threshold of $> 4.5$~K~km~$\textrm{s}^{-1}$ (cyan thick-line contours in Figure~\ref{fig:Fig1_TSmap}).
We assume the gamma-ray spectra of the cloud regions and HB9 SNR shell to  follow a simple power-law function of $dN/dE \propto E^{-{\rm \Gamma}}$.
Here, (i) the background model (the null hypothesis) consists of the radio template for the HB9 SNR shell as well as the Galactic and extragalactic diffuse emissions and 4FGL sources~\citep{Fermi-LAT:2019yla}.
To compare spatial models, we use the Akaike information criterion~(AIC; \citet{AIC1974ITAC}) defined as $2k - 2\log(L)$, where $k$ is the number of the estimated parameters in the model.
Consequently, we find that the model with the smaller AIC value is favored and that the better model gives a larger $\Delta{\rm AIC}~( = ({\rm AIC})_{\rm 0}-({\rm AIC})_{\rm m})$, where $({\rm AIC})_{\rm 0}$ and $({\rm AIC})_{\rm m}$ are the AIC values for the null hypothesis and a model including additional sources, respectively.
When we apply the CO template models to the two cloud regions (ii), $\Delta{\rm AIC}$ is found to be 44.4, indicating a significant correlation between the gamma-ray and $^{12}$ CO ($J=1-0$) line emissions (Table~\ref{tab:likelihood}).
Since the gamma-ray emission from R1 spatially coincides with PS J0506.5+4546 reported by~\citet{Sezer:2019fue}, we also apply a point source model to the same position as PS J0506.5+4546, instead of the CO template model of R1 (iii).
The resultant AIC is marginally ($1.3\sigma$ level) improved from the case with the CO template.
As a result, the gamma-ray emission from R1 is consistent with PS J0506.5+4546, while R2 is newly detected with a statistical significance of $6.1\sigma$ in this study.
%%%%%%%%%%%%%%%%%%%%%%%%%%%%%%%%%%%%%

%\begin{table*}[!t]
\begin{table*} 
    \tbl{Likelihood test results for spatial models and resultant spectral parameters of HB9 SNR shell and the two cloud regions, using the energy range of 1--500~GeV.
    The following three models are considered: (i) the background model (null hypothesis); (ii) additionally including gamma-ray emission from R1 and R2 estimated by the CO template model (see text); and (iii) assuming a point-source model, where the position of the source is the same as PS J0506.5+4546~\citep{Sezer:2019fue}, for R1 instead of the CO template model, while the CO template as in (ii) for R2.
    All sources of interest are fitted with a simple power-law function of $dN/dE \propto E^{-{\rm \Gamma}}$.
    The ``Flux'' and ``$\sqrt{\rm {TS}}$'' columns refer to the integral flux in the energy band of 1--500~GeV  in $10^{-10} {\rm cm^{-2}s^{-1}}$ and the statistical significance ($\sigma$) of each source approximated as $\sqrt{\mathrm{TS}}$, respectively.
    }{
    %\centering
    \begin{tabular}{cccccccccccc}
    \hline
    \multirow{2}{*}{Model} 
    & \multirow{2}{*}{$\Delta$AIC}
    & \multicolumn{3}{c}{\underbar{~R1~}} 
    & \multicolumn{3}{c}{\underbar{~R2~}} 
    & \multicolumn{3}{c}{\underbar{~HB9 SNR itself~}} \\
    & & Flux & $\Gamma$ & $\sqrt{\rm {TS}}$ & Flux & $\Gamma$ & $\sqrt{\rm {TS}}$ & Flux & $\Gamma$ & $\sqrt{\rm {TS}}$
    \\ \hline
    (i) & 0 & - & - & - & - & - & - & $24.9 \pm 2.1$ & $2.54 \pm 0.06$ & 13.2 \\
    (ii) & 44.4 & $2.0 \pm 0.8$ & $1.84 \pm 0.18$ & 4.5 & $4.8 \pm 1.2$ & $1.84 \pm 0.14$ & 6.1 & $24.3 \pm 2.1$ & $2.55 \pm 0.10$ & 12.8 \\
    (iii) & 58.2 & $1.3 \pm 0.5$ & $1.77 \pm 0.18$ & 5.8 & $4.8 \pm 1.2$ & $1.84 \pm 0.14$ & 6.1 & $24.5 \pm 2.1$ & $2.54 \pm 0.10$ & 13.0 \\
    \hline
    \end{tabular}}
    %\begin{tabnote}
    %\footnotemark[a] The background model (the null hypothesis). \\
    %\footnotemark[b] The gamma-ray emissions from R1 and R2 estimated with the CO template model (see text) are additionally included. \\
    %\footnotemark[c] A point-like source model, where the positions of the source  is the same as PS J0506.5+4546~\citep{Sezer:2019fue}, is assumed for R1, instead of the CO template model. For R2, the CO template is used as with Model (ii).\\
    %\end{tabnote}
    \label{tab:likelihood}
\end{table*}
%%%%%%%%%%%%%%%%%%%%%%%%%%%%%%%%%%%%%

\subsection{Spectral results}
\label{section:FermiAna_SED}

\begin{figure*}
    \begin{center}
    \includegraphics[width=15cm]{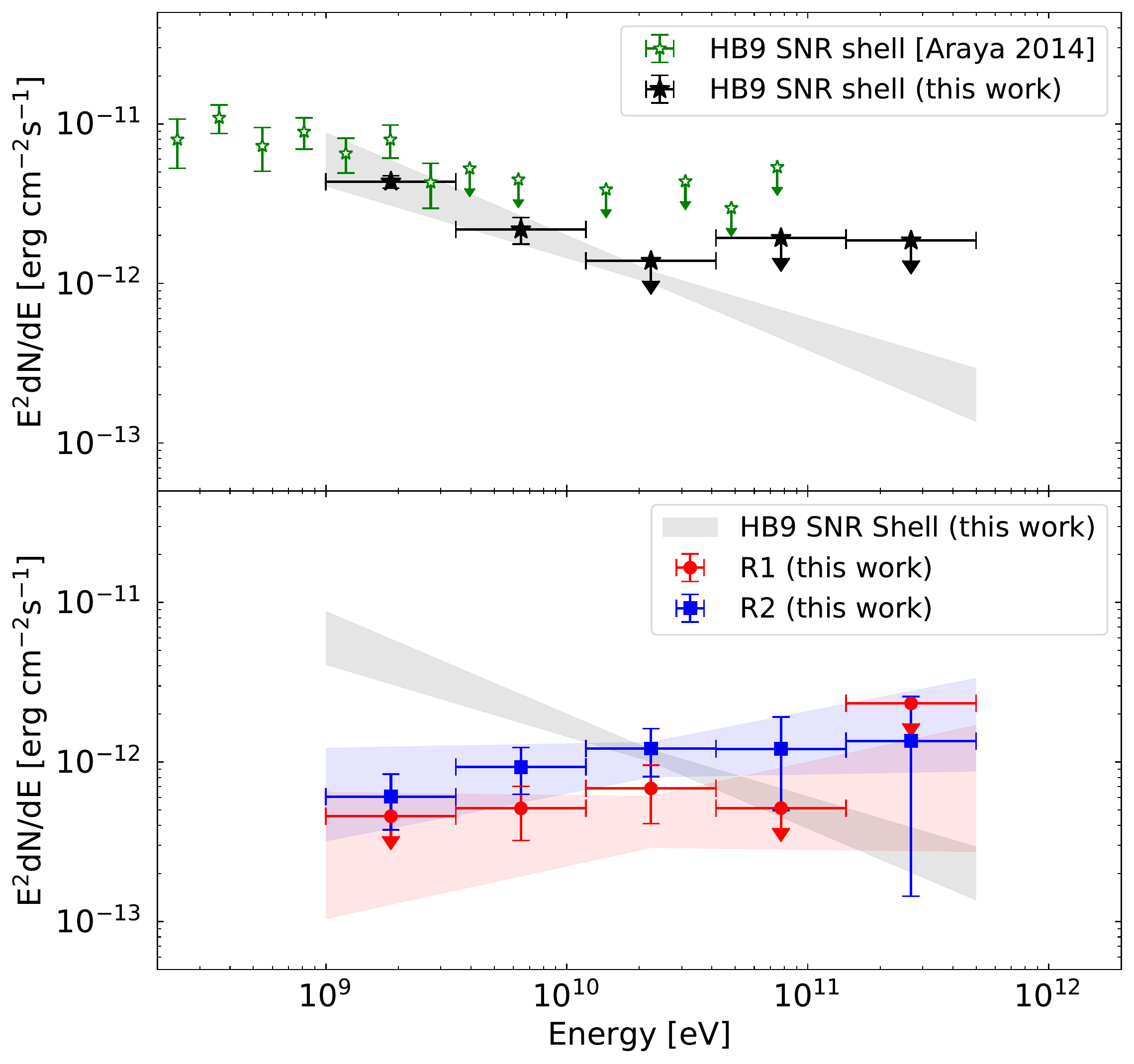}
    \end{center}
    \caption{
    Spectral energy distributions measured with the \textit{Fermi}-LAT (data points) and fit results in the gamma-ray band.
    {\bf Top}: Black and open green stars show the data of the SNR shell obtained in this work and \citet{Araya:2014kra}, respectively.
    {\bf Bottom}: Red and blue data represent the spectra of cloud regions R1 and R2, respectively. 
    Also shown as the shaded black region is the fit result of the SNR shell.
    % The lower panel shows the ratio between the flux of R2 and that of SNR shell and R1 (filled markers).
    % Solid lines in the respective colors show the model gamma-ray emissions derived on the basis of the hadronic scenario with the model parameters tabulated in Table~\ref{tab:modelpara}.
    }
    \label{fig:Fig2_SED_ModelBest}
\end{figure*}

We extract \textit{Fermi}-LAT energy spectra from the radio SNR shell region and two regions R1 and R2, individually, using the CO template model for the cloud regions.
Figure~\ref{fig:Fig2_SED_ModelBest} shows the resultant spectra for an energy range between 1 and 500~GeV.
%We find that the obtained \textit{Fermi}-LAT spectrum of the SNR shell is in good agreement with the gamma-ray spectrum reported by~\citet{Araya:2014kra} (Figure~\ref{fig:Fig2_SED_ModelBest}), the energy range of which is overall lower than that of  the former with some overlap.
% p7 l43 : "energy range of which is overall lower than..."   Not only difference, but what is new in the new analysis? Same as the reviewer's comment, is there any reason that the new analysis does not cover below 1GeV? New detection between 3GeV to 10GeV (?) and a stringent upper limit above 10GeV may be important for spectrum discussion later.     
We find that the obtained \textit{Fermi}-LAT spectrum of the SNR shell is consistent with the gamma-ray spectrum reported by~\citet{Araya:2014kra} (Figure~\ref{fig:Fig2_SED_ModelBest}). 
The respective best-fit parameters are summarized in Table~\ref{tab:likelihood}.
Here, a potential concern about the spectrum of R1 is that our assumption of diffuse gamma-ray spatial distribution may not be appropriate, given the spatial proximity with the already identified point source PS J0506.5+4546 (Figure~\ref{fig:Fig1_TSmap}). 
However, the discrepancy of the determined spectral properties (flux and index) between the results of the two spatial-distribution models is smaller than the  $1\sigma$ uncertainty (Table~\ref{tab:likelihood}). 
Therefore, we conclude that the difference in the results due to the difference between the assumed spatial distributions is not significant.
We note that the power-law index of R1 is consistent with that of PS J0506.5+4546 reported by~\citet{Sezer:2019fue}.
% The spectra are reproduced well by the model which will be discussed in Section~\ref{section:Discussion}.

%\section{Discussion} \label{section:Discussion}
\section{Spectral modeling and its implication} \label{section:Discussion}

In this work, we explore the possibility that the R1 and R2 are attributed to the delayed gamma-ray from MCs illuminated by the CRs accelerated in HB9.
%In the following section, we calculate the delayed gamma-ray spectra in the MC regions, R1 and R2, using the method described in Section~\ref{section:Model}.
In the following section, we calculate the delayed gamma-ray spectra in the MC regions, R1 and R2, using the method described in Section~\ref{section:Model}, and try to fit the observed spectra of the cloud regions and the SNR shell simultaneously.
According to \citet{Araya:2014kra}, once the gamma-ray spectrum of the SNR shell is modeled with the hadronic emission, it would require enormous explosion energy of supernova or a dense density of the interstellar medium (ISM)\footnote{\citet{2021arXiv211208748E} explored the possibility of the hadronic origin for the gamma-ray from the HB9 shell using 10-year data of Fermi-LAT nevertheless found no hint. However, since they performed the modeling study without radio and X-ray data, they could not restrict the electron distribution from synchrotron radiation and thus could not judge whether the hadronic or leptonic model is preferred.}.
Hence, we assume that the gamma-ray emission in the SNR shell originates mainly from the leptonic processes.
On the other hand, we consider two cases for the origin of gamma-ray emissions from the cloud regions: one where the leptonic emission dominates (Leptonic-dominated model), and the other where the hadronic emission dominates (Hadronic-dominated model) in the GeV band.
We present the results of the modeling in Section \ref{section:Modeling2} and discuss implications on the model parameters in Section \ref{section:Modeling3}.

\subsection{Modeling delayed gamma-ray emission} \label{section:Model}

The energy spectra of the gamma-ray emissions from the MCs around SNR HB9 are calculated following the method proposed by \citet{2009MNRAS.396.1629G} and \citet{2011MNRAS.410.1577O}. 
%\oldwiadd{Ohira+11} (\oldwicom{https://ui.adsabs.harvard.edu/abs/2011MNRAS.410.1577O/abstract}).
They considered the escape process of the CR protons, which are accelerated by an SNR, from a shock front into interstellar space during the Sedov phase.
Escaped CR protons emit high-energy gamma-rays via $\pi^{0}$ production through p-p collision in MCs in the vicinity of the SNR.
In addition, we also discuss the contribution from inverse Compton scattering and non-thermal bremsstrahlung from CR electrons escaping from the SNR.
We evaluate later in this section the expected gamma-ray fluxes from the R1 and R2 regions in conjunction with the energy spectra of the escaped CR protons and electrons.

The radio continuum contours (Figure~\ref{fig:Fig1_TSmap}) show a circular symmetric morphology of SNR HB9, suggesting it has probably maintained a spherically symmetric structure throughout its evolution.
Therefore, we assume that the SNR shell itself and the escaped CR distribution are spherically symmetric.
%The hydrogen density $n_{\rm ISM}$ of the interstellar medium (ISM) in SNR HB9 is $n_{\rm ISM}=0.2 ~{\rm cm}^{-3}$~\citep{Leahy:2006vs}.
Based on X-ray observations, the energy of the supernova explosion ($E_{\rm SN}$) was estimated to be (0.15--0.30)$\times10^{51}$~erg, and the relation between the explosion energy and the hydrogen density of ISM ($n_{\rm ISM}$) was also estimated as $(E_{\rm SN}/10^{51}~\rm{erg}) (n_{\rm ISM}/1~{\rm cm}^{-3})^{-1} = 5$~\citep{Leahy:2006vs}. In this paper, we adopt $0.3\times10^{51}$~erg as the explosion energy, in which case the density is given as $0.06~{\rm cm}^{-3}$.
If the initial velocity of the blast wave was $u_{\rm sh,0}=10^9~{\rm cm}~{\rm s}^{-1}$, HB9 entered the Sedov phase at the age $t_{\rm Sedov}=3.6\times10^{2}~{\rm yr}$ when its radius  was $3.6~{\rm pc}$.
%If the energy of the supernova explosion was $E_{\rm SN}=10^{51} ~{\rm erg}$ and if the initial velocity of the blast wave was $u_{\rm sh,0}=10^9~{\rm cm}~{\rm s}^{-1}$, HB9 entered the Sedov phase at the age $t_{\rm Sedov}=3.6\times10^{2}~{\rm yr}$  when its radius  was $3.6~{\rm pc}$.
Given that the observed radius of HB9 is $\sim 10~{\rm pc}$, which is larger than 3.6~pc, HB9 must be already in the Sedov phase.
\citet{Leahy:2006vs} estimated the Sedov age of SNR HB9 to be $6.6\times10^{3}$~yr, which we adopt as the age of the SNR in this discussion.

First of all, let us model the energy spectrum of CR protons.
During the Sedov phase, the maximum energy of the CR protons accelerated at the SNR shock is determined by the timescale of escape from the acceleration region, and decreases with time \citep{2005A&A...429..755P,2009MNRAS.396.2065C,2010A&A...513A..17O}.
The temporal evolution of the maximum energy depends on non-linear processes, such as amplifying the magnetic field at the shock, which is still theoretically unclear.
Thus, we adopt the phenomenological power-law dependence of the cutoff energy $E_{\rm esc}$ of the proton spectrum at the SNR shock on the age $t$ of the SNR, as discussed by \citet{2009MNRAS.396.1629G,2010A&A...513A..17O}:
%%%%%%%%%%%%%%%%%%%%%%%%%%%%%%%%%%%%%
\begin{equation}\label{eq:Ecut}
E_{\rm esc}(t)=E_{\max}\left(\frac{t}{t_{\rm Sedov}}\right)^{-\alpha},
\end{equation}
%%%%%%%%%%%%%%%%%%%%%%%%%%%%%%%%%%%%%
where $E_{\max}$ is the maximum energy of the CR protons at $t_{\rm Sedov}$ and is set to $3$~PeV, and $\alpha$ is the power-law index that is determined so that the current maximum energy ($E_{\rm now}$) is equal to $E_{\rm esc}(t)$ at the current age $t_{\rm age}=6.6\times10^{3}$~yr~\citep{Leahy:2006vs}.
Here, we adopt $E_{\rm now}=300~{\rm GeV}$, which is the same value as that of the electrons at the SNR shell at present (see also Section \ref{section:Modeling2}).
However, in general, the maximum energy of CRs in the SNR shell depends on the particle species and is determined by the balance between acceleration and escape from the shell or energy loss due to radiative cooling.
\citet{Araya:2014kra}, assuming that radiative cooling limits the maximum CR electron energy, estimated that the gyro factor would be a too large value of $\sim$ 660 compared to the standard SNR value.
In such a case, as shown in \citet{Ohira:2011xq}, the escape process determines both the highest energies of CR electrons and protons.
Therefore, according to \citet{Ohira:2011xq}, we can assume that the current maximum energy of electrons is the same as that of the protons.
The timescale for a particle with the energy $E$, $t_{\rm esc}(E)$, to escape into interstellar space from the supernova explosion, as a function of the particle energy $E$ is given by      
%The timescale for a particle with the energy $E$ to escape into interstellar space, or the time $t_{\rm esc}(E)$ for it from the supernova explosion, as a function of the particle energy $E$ is given by
%%%%%%%%%%%%%%%%%%%%%%%%%%%%%%%%%%%%%
\begin{equation}\label{eq:t_esc}
t_{\rm esc}(E)=t_{\rm Sedov}\left(\frac{E}{E_{\max}}\right)^{-1/\alpha}.
\end{equation}
%%%%%%%%%%%%%%%%%%%%%%%%%%%%%%%%%%%%%

The distribution function $f_{p,\rm out}$ per unit energy per unit volume of the CR protons escaping into interstellar space can be obtained by solving the transport equation,
%%%%%%%%%%%%%%%%%%%%%%%%%%%%%%%%%%%%%
\begin{equation}\label{eq:transport}
    \frac{\partial f_{p,\rm out}}{\partial t}(t, r, E)-D_{\mathrm{ISM}}(E) \Delta f_{p,\rm out}(t, r, E)=q_{p,s}(t, r, E),
\end{equation}
%%%%%%%%%%%%%%%%%%%%%%%%%%%%%%%%%%%%%
where $t$ is the age of the SNR, $r$ is the distance from the SNR center, $D_{\mathrm{ISM}}$ is the diffusion coefficient in the ISM, and $q_{p,s}$ is the injection rate of the CR protons from the SNR shock into interstellar space per unit energy, unit volume, and unit time.
We adopt the following form of the diffusion coefficient $D_{\rm ISM}$:
%%%%%%%%%%%%%%%%%%%%%%%%%%%%%%%%%%%%%
\begin{equation}\label{eq:DISM}
D_{\mathrm{ISM}}(E)=D_0 \left(\frac{E}{10~{\rm GeV}}\right)^{\delta},
\end{equation}
%%%%%%%%%%%%%%%%%%%%%%%%%%%%%%%%%%%%%
where $D_0$ is the diffusion coefficient of CRs at $E=10 ~{\rm GeV}$. We assume $D_0=3 \times 10^{28} ~{\rm cm}^2~{\rm s}^{-1}$ and the index $\delta=1/3$, the latter of which is consistent with the Galactic mean expected in the CR propagation model~(e.g., \citet{Blasi:2011fm}).
%For simplicity, following \citet{2009MNRAS.396.1629G}, we assume that CRs are injected into the center of the SNR.
% ~Ohira+11\oldwicom{Reference:https://ui.adsabs.harvard.edu/abs/2011MNRAS.410.1577O/abstract}
Following~\citet{2011MNRAS.410.1577O}, we assume that CRs with the energy $E$ are injected from the SNR shell $R_{\rm esc}=u_{\rm sh,0}t_{\rm Sedov}(t_{\rm esc}(E)/t_{\rm Sedov})^{2/5}$ at $t=t_{\rm esc}(E)$.
We also assume that the energy spectrum of the injected CRs is monochromatic with the energy $E$ and that particles start to escape at any given time (see equation (\ref{eq:t_esc})).
Taking account of these assumptions, the injection rate $q_{\rm s}$ is given by
%%%%%%%%%%%%%%%%%%%%%%%%%%%%%%%%%%%%%
\begin{equation}\label{eq:qs}
    % \oldwidel{q_s=N_{\mathrm{esc}}(E) \delta(\boldsymbol{r}) \delta\left(t-t_{\rm esc}(E)\right),\\}
    q_{p,s}=\frac{N_{\mathrm{esc}}(E)}{4\pi r^2} \delta(r-R_{\rm esc}(E)) \delta\left(t-t_{\rm esc}(E)\right),
\end{equation}
%%%%%%%%%%%%%%%%%%%%%%%%%%%%%%%%%%%%%
where $r$ is the displacement from the center of the SNR, $N_{\rm esc}$ is the spectrum of all the CRs that have escaped from the SNR up to the present time.
In this paper, we assume that the total energy of the escaped CR protons is proportional to the explosion energy of the supernova explosion $E_{\rm SN}$, namely
%%%%%%%%%%%%%%%%%%%%%%%%%%%%%%%%%%%%%
\begin{equation}\label{eq:Nesc_normalize}
    \int_{E_{\rm now}}^{E_{\max}} E N_{\rm esc}(E) dE =\eta E_{\rm SN},
\end{equation}
%%%%%%%%%%%%%%%%%%%%%%%%%%%%%%%%%%%%%
where $\eta$ is the acceleration efficiency coefficient.
%\widel{Using the $E^{-2}$ dependence of the CR spectrum (which is the same as the electron index at HB9 as mentioned in Section~\ref{sec:intro}), the integral of the equation (\ref{eq:Nesc_normalize}) yields, derived in~\citet{Araya:2014kra}}
Here, we assume that $N_{\rm esc}$ is a power-law function of $E$ with an index $p_{\rm esc}$.
%\textbf{Ohira+10}
According to~\citet{2010A&A...513A..17O}, it is shown that the index $p_{\rm esc}$ of CRs escaping from the SNR is steeper than the index $p_{\rm SNR}$ of CRs  confined in the SNR shell.
While the value of $p_{\rm esc}$ is determined by the time evolution of the maximum energy of CRs at the SNR shell and the CR production rate, we treat $p_{\rm esc}$ as a parameter.
By assuming a power-law form of $N_{\rm esc}$, the integral of the equation (\ref{eq:Nesc_normalize}) yields,
% %%%%%%%%%%%%%%%%%%%%%%%%%%%%%%%%%%%%%
% \begin{equation} \label{eq:Nesc}
% N_{\rm esc}(E)=\frac{\eta E_{\rm SN}}{\ln\left(E_{\max}/E_{\rm now}\right)}E^{-2}.
% \end{equation}
% %%%%%%%%%%%%%%%%%%%%%%%%%%%%%%%%%%%%%
%%%%%%%%%%%%%%%%%%%%%%%%%%%%%%%%%%%%%
%\begin{equation} \label{eq:Nesc}
\begin{eqnarray} \label{eq:Nesc}
N_{\rm esc}(E)=
\left\{
    \begin{array}{ll}
    \frac{\eta \left(2-p_{\rm esc}\right) E_{\rm SN}}{E_{\rm now}^2}
    %\left[\left(\frac{E_{\rm max}}{E_{\rm now}}\right)^{2-p_{\rm esc}}-1\right]^{-1}\left(\frac{E}{E_{\rm now}}\right)^{-p_{\rm esc}}
    % & p_{\rm esc}\neq 2\\
    \\ ~\times \left[\left(\frac{E_{\rm max}}{E_{\rm now}}\right)^{2-p_{\rm esc}}-1\right]^{-1} \\ ~\times \left(\frac{E}{E_{\rm now}}\right)^{-p_{\rm esc}} & p_{\rm esc}\neq 2\\
    \frac{\eta E_{\rm SN}}{\ln\left(E_{\max}/E_{\rm now}\right)}E^{-2} & p_{\rm esc}=2
    \end{array}
    \right.
    .
%\end{equation}
\end{eqnarray}
%%%%%%%%%%%%%%%%%%%%%%%%%%%%%%%%%%%%%
The propagation models~\citep{2012ApJ...752...69O, 2017PhRvD..95h3007Y} expect the spectral index of particles escaped from Galactic CR origin to be $\sim2.4$, which is adopted as a fiducial parameter of $p_{\rm esc}$ in this modeling. The effect of this parameter on the modeling will be discussed in Section~\ref{section:Modeling3}.
The energy spectrum of the escaped CRs is obtained by combining the transport equation (\ref{eq:transport}) and equations (\ref{eq:qs}) and (\ref{eq:Nesc}).
Specifically, the solution of equation (\ref{eq:transport}) for $E>E_{\rm esc}(t)$ is given by, according to~\citet{2011MNRAS.410.1577O} 
%\oldwidel{\citet{2009MNRAS.396.1629G}} Ohira+11b\oldwicom{Reference:https://ui.adsabs.harvard.edu/abs/2011MNRAS.410.1577O/abstract}, % Original: \citep{2009MNRAS.396.1629G}:
%%%%%%%%%%%%%%%%%%%%%%%%%%%%%%%%%%%%%
%\begin{equation}\label{eq:f_out}
\begin{eqnarray} \label{eq:f_out}
%f_{\rm{p,out }}(t, r, E)&=&\frac{N_{\rm {esc }}(E)}{4 \pi^{3 / 2} r R_{\rm {esc}} R_{\mathrm{d,p}}}\left[\exp \left(-\frac{\left(r-R_{\rm{esc}}\right)^{2}}{R_{\rm d,p}^{2}}\right)-\exp \left(-\frac{\left(r+R_{\rm{esc }}\right)^{2}}{R_{\rm d,p}^{2}}\right)\right]
f_{\rm{p,out }}(t, r, E) &=& \frac{N_{\rm {esc }}(E)}{4 \pi^{3 / 2} r R_{\rm {esc}} R_{\mathrm{d,p}}} \nonumber \\ && \times \left[\exp \left(-\frac{\left(r-R_{\rm{esc}}\right)^{2}}{R_{\rm d,p}^{2}}\right)-\exp \left(-\frac{\left(r+R_{\rm{esc }}\right)^{2}}{R_{\rm d,p}^{2}}\right)\right]
%\end{equation}
\end{eqnarray}
%%%%%%%%%%%%%%%%%%%%%%%%%%%%%%%%%%%%%
where $R_{\rm d,p}$ is the diffusion length of a CR proton with the energy $E$, defined as
%%%%%%%%%%%%%%%%%%%%%%%%%%%%%%%%%%%%%
\begin{equation}\label{eq:RDiff}
R_{\rm d,p}\equiv\sqrt{4D_{\mathrm{ISM}}(E)\left(t-t_{\rm esc}(E)\right)}.
\end{equation}
%%%%%%%%%%%%%%%%%%%%%%%%%%%%%%%%%%%%%

In order to obtain the energy spectrum of CR electrons $f_{e,{\rm out}}$, it is required to solve the following transport equation with radiative cooling:
%%%%%%%%%%%%%%%%%%%%%%%%%%%%%%%%%%%%%
%\begin{equation}\label{eq:transport_e}
\begin{eqnarray} \label{eq:transport_e}
    \frac{\partial f_{e,\rm out}}{\partial t}(t, r, E)
    && - D_{\mathrm{ISM}}(E) \Delta f_{e,\rm out}(t, r, E) \nonumber \\
    && +\frac{\partial}{\partial E}\left(P(E)f_{e,\rm out}(t, r, E)\right)
    %=q_{e,s}(t, r, E),
    = q_{e,s}(t, r, E),
%\end{equation}
\end{eqnarray}
%%%%%%%%%%%%%%%%%%%%%%%%%%%%%%%%%%%%%
where $P(E)$ is an energy loss rate of CR electrons, and $q_{e,s}$ is an injection rate of CR electrons, which can be written as
%%%%%%%%%%%%%%%%%%%%%%%%%%%%%%%%%%%%%
\begin{equation}\label{eq:qse}
    q_{e,s}=K_{\rm ep}q_{p,s},
\end{equation}
%%%%%%%%%%%%%%%%%%%%%%%%%%%%%%%%%%%%%
where $K_{\rm ep}$ is the ratio of the total energy of CR electrons to CR protons.
As a cooling process, we consider only the synchrotron radiation, which is the most dominant effect for electrons with energies above $\mathcal{O}({\rm GeV})$ (e.g., \citet{1990acr..book.....B}). %\oldwicom{https://ui.adsabs.harvard.edu/abs/1990acr..book.....B/abstract}.

The temporal evolution of the maximum energy of CR electrons in the SNR shell is different from that of protons due to radiative cooling.
After entering the Sedov phase, the maximum energy of CR electrons is first determined by the balance between the radiative cooling and acceleration, and, after that, by the balance between acceleration and escape from the SNR shell, similar to the proton~\citep{Ohira:2011xq}.
When these two phases switch, CR electrons start to escape from the SNR shell and be injected into the interstellar space.
In this paper, we parametrize this switching time $t_e$ as follows:
%%%%%%%%%%%%%%%%%%%%%%%%%%%%%%%%%%%%%
\begin{equation}
    t_e=\xi_e t_{\rm Sedov},
\end{equation}
%%%%%%%%%%%%%%%%%%%%%%%%%%%%%%%%%%%%%
where $\xi_e$ is the time in the unit of $t_{\rm Sedov}$ at which the electron starts to escape, and in the limit of $\xi_e\to1$, the injection is identical to that of the proton (see equation (\ref{eq:qs})).
After $t_e$, since the time evolution of the maximum energy of electrons is expected same as that of the proton~\citep{Ohira:2011xq}, we use equations (\ref{eq:Ecut}) and (\ref{eq:t_esc}).
Originally, $\xi_e$ is given based on the assumed environment~\citep{Ohira:2011xq}, but we set $\xi_e = 1$ to consider the limit where the gamma-ray flux of leptonic emissions is maximized in this modeling.
The solution to equation (\ref{eq:transport_e}) can be derived using the method described in Appendix \ref{Appendix:derivation}.
%\widel{For \wiadd{$E_{\rm esc}(t)<E<E_{\rm esc}(t_e)$}, the solution can be written as:}

Considering only the synchrotron radiation as the cooling process (i.e. $P(E)=Q_{\rm syn}E^2$, see equation (\ref{eq:syn_cooling}) of Appendix \ref{Appendix:derivation} for details), the solution for $E_{\rm esc}(t)<E<E_{\rm esc}(t_e)$ can be written as:
%%%%%%%%%%%%%%%%%%%%%%%%%%%%%%%%%%%%%
%\begin{equation}\label{eq:f_oute}
\begin{eqnarray}\label{eq:f_oute}
%f_{\rm {e,out}}(t, r, E)&=&\frac{K_{\rm ep}N_{\mathrm{esc}}\left(E_{c}\right)}{4 \pi^{3 / 2} r R_{\mathrm{c}} R_{\rm d, e}} \frac{E_{c}^{2}}{E^{2}} \frac{1}{1-Q_{\rm syn}t_{c}E_c / \alpha} \left[\exp \left(-\frac{\left(r-R_{\mathrm{c}}\right)^{2}}{R_{\rm d, e}^{2}}\right)-\exp \left(-\frac{\left(r+R_{\mathrm{c}}\right)^{2}}{R_{\rm d, e}^{2}}\right)\right].
&& f_{\rm {e,out}}(t, r, E) = \frac{K_{\rm ep}N_{\mathrm{esc}}\left(E_{c}\right)}{4 \pi^{3 / 2} r R_{\mathrm{c}} R_{\rm d, e}} \frac{E_{c}^{2}}{E^{2}} \frac{1}{1-Q_{\rm syn}t_{c}E_c / \alpha} \nonumber \\ && \qquad\quad \times \left[\exp \left(-\frac{\left(r-R_{\mathrm{c}}\right)^{2}}{R_{\rm d, e}^{2}}\right)-\exp \left(-\frac{\left(r+R_{\mathrm{c}}\right)^{2}}{R_{\rm d, e}^{2}}\right)\right].
%\end{equation}
\end{eqnarray}
%%%%%%%%%%%%%%%%%%%%%%%%%%%%%%%%%%%%%
% where $t_c$ and $E_c$ are functions of $E$ and $t$ defined through a solution to the algebraic equation (\ref{eq:tceqs}) (see Appendix \ref{Appendix:derivation} for details).

Once the energy spectrum of the CRs has been obtained, the flux of the gamma-ray emission is calculated based on the neutral pion decay process in MCs for CR protons, and inverse Compton scattering and the relativistic bremsstrahlung for CR electrons.
Here we assume that the MC is ``optically'' thin for the CRs and that the cloud consists of a spatially uniform gas with the total mass $M_{\rm cl}$.
% For simplicity, we calculate the gamma-ray spectra using the CR density at the center of the MC.
% This simplification is justified because the spatial distribution of the CR density in the MC has negligible effect on the value of the integrated gamma-ray flux for the sizes of the MCs of our interest, R1 and R2.
For simplicity, we also assume that the MC is a sphere of radius $R_{\rm cl}$.
The spectrum of CRs in the MC can be written as follows:
%%%%%%%%%%%%%%%%%%%%%%%%%%%%%%%%%%%%%
\begin{equation}
    N_{\rm cl}(E)=\int_{d_{\mathrm{cl}}-R_{\mathrm{cl}}}^{d_{\mathrm{cl}}+R_{\mathrm{cl}}} \frac{\pi r}{d_{\mathrm{cl}}}\left(R_{\mathrm{cl}}^{2}-\left(r-d_{\mathrm{cl}}\right)^{2}\right) f_{s,\mathrm{out}}(r) d r,
\end{equation}
%%%%%%%%%%%%%%%%%%%%%%%%%%%%%%%%%%%%%
where $d_{\rm cl}$ is a distance to the center of the MC from the center of the SNR, and $s$ is species ($\rm e$=electron, $\rm p$=proton).
Note that, for the parameters used in this paper, a spatial gradient of the CR distribution is not important, and only a few \% difference occurs even if $N_{\rm cl} \sim 4\pi R_{\rm cl}^3f_{s,{\rm out}}(d_{\rm cl})/3$ is used.
% $dN_{\rm cl}/dE \sim 4\pi R_{\rm cl}^3f_{s,{\rm out}}(d_{\rm cl})/3$

The energy spectrum of CRs in the SNR shell, $N_{s,{\rm shell}}$, is calculated consistently with the distribution function of escaped particles.
By using the normalization equation (\ref{eq:Nesc_normalize}) same as equation (\ref{eq:Nesc}), the spectra at $E<E_{\rm esc}$ of CR protons and electrons in the SNR shell can be written as
% %%%%%%%%%%%%%%%%%%%%%%%%%%%%%%%%%%%%%
% \begin{equation}
% N_{p,{\rm shell}}(E)=\frac{\eta E_{\rm SN}}{\ln\left(E_{\max}/E_{\rm now}\right)}E^{-2},
% \end{equation}
% %%%%%%%%%%%%%%%%%%%%%%%%%%%%%%%%%%%%%
%%%%%%%%%%%%%%%%%%%%%%%%%%%%%%%%%%%%%
\begin{equation}
N_{p,{\rm shell}}(E)=N_{\rm esc}(E_{\rm now})\left(\frac{E}{E_{\rm now}}\right)^{-p_{\rm SNR}},
\end{equation}
%%%%%%%%%%%%%%%%%%%%%%%%%%%%%%%%%%%%%
and
% %%%%%%%%%%%%%%%%%%%%%%%%%%%%%%%%%%%%%
% \begin{equation}
% N_{e,{\rm shell}}(E)=K_{\rm ep}\frac{\eta E_{\rm SN}}{\ln\left(E_{\max}/E_{\rm now}\right)}E^{-2},
% \end{equation}
% %%%%%%%%%%%%%%%%%%%%%%%%%%%%%%%%%%%%%
%%%%%%%%%%%%%%%%%%%%%%%%%%%%%%%%%%%%%
\begin{equation}
N_{e,{\rm shell}}(E)=K_{\rm ep}N_{\rm esc}(E_{\rm now})\left(\frac{E}{E_{\rm now}}\right)^{-p_{\rm SNR}},
\end{equation}
%%%%%%%%%%%%%%%%%%%%%%%%%%%%%%%%%%%%%
respectively,
where $p_{\rm SNR}$ is the index of CRs that have not yet escaped from the SNR shell.
We fix $p_{\rm SNR}=2$ in this paper because the index $p_{\rm SNR}$ of HB9 has been well determined by observations of radio continuum (see Section~\ref{sec:intro}).

To calculate the spectra of non-thermal emissions, we use the radiative code \textit{naima} \citep{2015ICRC...34..922Z}.
%The parameterization of the neutral pion decay by \citet{2014PhRvD..90l3014K} is used to calculate the hadronic gamma-rays.
As for the seed photon fields in the IC process, we assume the cosmic microwave background and Galactic far-infrared (FIR) radiation, the latter of which was not considered in the modeling of \citet{Araya:2014kra}. 
The energy density of the FIR radiation is estimated to be 0.099~eV cm$^{-3}$ at $T$ = 27~K, using the package GALPROP \citep{Vladimirov:2010aq}.

\begin{table*}  
    \caption{Fiducial parameters used to calculate the model spectra} \label{tab:modelpara}
    \centering
    \begin{tabular}{lccc}
    \hline
    SNR parameters &  Symbol & & \\\hline
    SN explosion energy & $E_{\rm SN}$ & \multicolumn{2}{c}{$0.3\times10^{51}~{\rm erg}$}    \\ 
    Initial shock velocity & $u_{\rm sh}$ & \multicolumn{2}{c}{$10^{9}~{\rm cm}~{\rm s}^{-1}$}    \\ 
    Age of the SNR & $t_{\rm age}$ & \multicolumn{2}{c}{$6.6\times10^{3}~{\rm yr}$}    \\ 
    Distance to the SNR &  & \multicolumn{2}{c}{$0.8~{\rm kpc}$}    \\ 
    Acceleration efficiency\footnotemark[$\dagger$] & $\eta$ & \multicolumn{2}{c}{$0.1$ (0.003)}    \\ 
    Electron to proton flux ratio\footnotemark[$\dagger$] & $K_{\rm ep}$ & \multicolumn{2}{c}{$0.02$ (1)} \\
    Maximum CR energy at $t=t_{\rm Sedov}$ & $E_{\max}$ & \multicolumn{2}{c}{$3~{\rm PeV}$}    \\ 
    Current maximum energy of CRs & $E_{\rm now}$ & \multicolumn{2}{c}{$300~{\rm GeV}$}    \\ 
    Magnetic field in the SNR & $B_{\rm SNR}$ & \multicolumn{2}{c}{$8~{\rm \mu G}$}    \\
    Particle index in the SNR & $p_{\rm SNR}$ & \multicolumn{2}{c}{2.0}    \\
    Particle index after escaping from SNR & $p_{\rm esc}$ & \multicolumn{2}{c}{2.4}    \\
    \hline ISM parameters &  Symbol & & \\\hline
    Number density & $n_{\rm ISM}$ & \multicolumn{2}{c}{$0.06~{\rm cm}^{-3}$}    \\ 
    Diffusion coefficient at $E=10~{\rm GeV}$ & $D_0$ & \multicolumn{2}{c}{$3 \times 10^{28}~{\rm cm}^2~{\rm s}^{-1}$}    \\ 
    Index of dependence on $E$ of diffusion & $\delta$ & \multicolumn{2}{c}{$1/3$}    \\ 
    Magnetic field in ISM & $B_{\rm ISM}$ & \multicolumn{2}{c}{$3~{\rm \mu G}$}    \\ 
    \hline  Molecular cloud (MC) parameters &  Symbol & R1 & R2 \\\hline
    Distance to the MC from SNR & $d_{\rm cl}$ & $17.8~{\rm pc}$ & $39.4~{\rm pc}$\\
    Radius of the MC & $R_{\rm cl}$ & $3.6~{\rm pc}$ & $7.0~{\rm pc}$\\
    Average hydrogen number density\footnotemark[$\dagger$] & $n_{\rm H}$ & $150~(1000)~{\rm cm}^{-3}$ & $200~(1000)~{\rm cm}^{-3}$ \\
    Mass of the  MC\footnotemark[$\dagger$] & $M_{\rm cl}$ & $730~(4800)~M_\odot$ & $7100~(36000)~M_\odot$\\
    %Energy density      & & $1.12~{\rm eV cm}^{-3}$ & $0.35~{\rm eV cm}^{-3}$ \\
    Reflected SNR age   & $t_{\rm ref}$ & $-1.7\times10^{2}~{\rm yr}$ & $-7.5\times10^{2}~{\rm yr}$ \\
    Magnetic field in MC & $B_{\rm MC}$ & \multicolumn{2}{c}{$3~{\rm \mu G}$}    \\ 
    \hline
    \end{tabular}
    \begin{tabnote}
    \footnotemark[$\dagger$] The values in parentheses are used in the leptonic-dominated model. \\
    \end{tabnote}
\end{table*}
%%%%%%%%%%%%%%%%%%%%%%%%%%%%%%%%%%%%%

\subsection{Application to the cloud regions around HB9} \label{section:Modeling2}

In order to calculate the gamma-ray flux at the cloud regions, R1 and R2, the distance between the MCs and SNR is required.
By fitting the CO intensity map integrated over the velocity range between $-10.4$ and $+2.6$~km s$^{-1}$ with a two-dimensional symmetric Gaussian, we determine the center positions of clouds R1 and R2 to be (l, b) = (161.8 $\pm$ 0.1$^{\circ}$, 2.8 $\pm$ 0.1$^{\circ}$) and (162.6 $\pm$ 0.1$^{\circ}$, 1.6 $\pm$ 0.1$^{\circ}$), respectively, in the galactic coordinates.
Then, the projected distances between the centers of the SNR and clouds R1 and R2 are calculated to be 17.8 and 39.4~pc, respectively, using the distance to the SNR from the Earth of 0.8~kpc (Section~\ref{sec:intro} and Table~\ref{tab:modelpara}).
We  treat these projected distances as the actual distances in our discussion, although  they should  be considered as the lower limit of the true distances in reality due to the uncertainty of their locations along the line of sight. 
We also obtain the radii of the MCs from the standard deviation of the fitted Gaussian, which are 3.6~pc and 7.0~pc for R1 and R2, respectively.

%We perform a broadband modeling of the non-thermal emissions from HB9 SNR shell and the two cloud regions.
For the spectral modeling, the data points in the radio band for the HB9 SNR shell are obtained from the literature~\citep{Dwarakanath:1982jap, Reich:2003aar, Leahy:2006vs, Roger:1999jy, 2011A&A...529A.159G}, while the radio flux of the mean local background including the cloud region at a frequency of 865~MHz~\citep{Reich:2003aar} is used as an upper limit for the cloud regions.
In the X-ray band, only the thermal emission from the hot gas inside the shell has been detected, while non-thermal emission has not been measured~(e.g, \citet{Leahy:1995aa, Sezer:2019fue}).
We adopt the 0.1--2.5~keV flux of the thermal emission from HB9~\citep{Leahy:1995aa} as an upper limit for the non-thermal emission in the energy band.

\begin{figure*}
    \begin{center}
    \includegraphics[width=16.5cm]{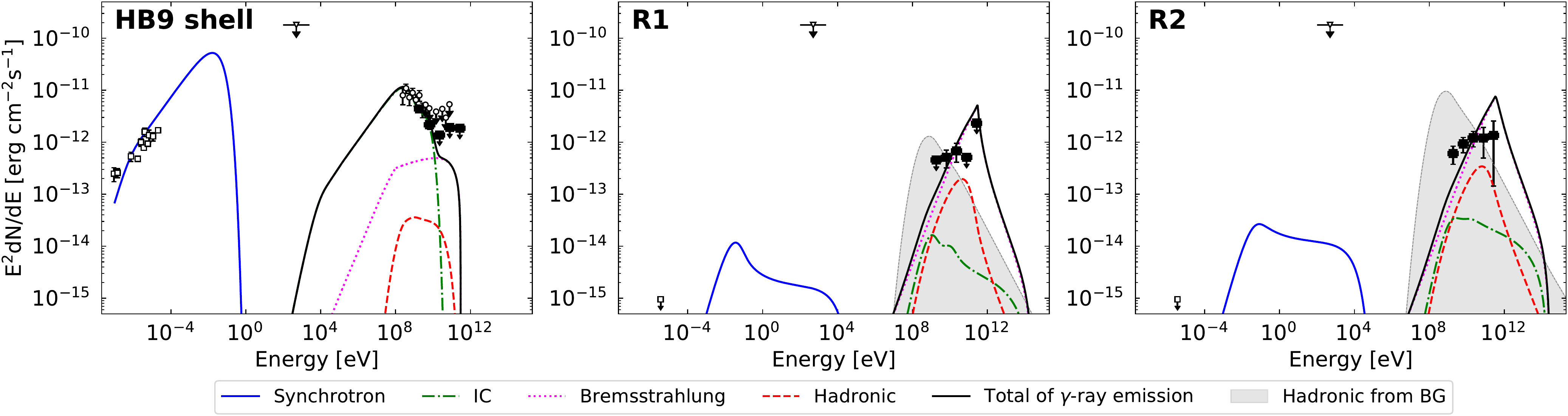}
    \end{center}
    \caption{
    Broadband spectral energy distributions of the non-thermal emission from the HB9 shell and the cloud regions with the leptonic-dominated model, using the parameters tabulated in Table~\ref{tab:modelpara}.
    The left, middle, and right panels show the results about the HB9 shell, R1 and R2, respectively.
    The radio and X-ray data are taken from~\citet{Dwarakanath:1982jap, Reich:2003aar, Leahy:2006vs, Roger:1999jy, 2011A&A...529A.159G} and~\citet{Leahy:1995aa}, respectively (see text for detail).
    The filled squares and open circles show the \textit{Fermi}-LAT data derived in Section \ref{section:FermiAna_SED} and~\citet{Araya:2014kra}, respectively.
    The lines represent each component of the emission models: synchrotron (blue), electron bremsstrahlung (magenta), IC (green), neutral pion decay (red), and the total gamma-ray emissions (black).
    In order to demonstrate how much of the Galactic diffuse background at the cloud regions, which may affect the systematic uncertainties in the \textit{Fermi} analysis, we show the hadronic emission from the background CRs with the energy spectrum $J_{\mathrm{CR}}(E)=2.2(E/\mathrm{GeV})^{-2.75} \mathrm{~cm}^{-2} \mathrm{~s}^{-1} \mathrm{GeV}^{-1} \mathrm{sr}^{-1}$ (e.g., \citet{1986A&A...157..223D}) as the shaded grey region.
    %Also shown as the shaded grey region is the hadronic emission from the background CRs with energy spectrum $J_{\mathrm{CR}}(E)=2.2(E/\mathrm{GeV})^{-2.75} \mathrm{~cm}^{-2} \mathrm{~s}^{-1} \mathrm{GeV}^{-1} \mathrm{sr}^{-1}$ (e.g., \citet{1986A&A...157..223D}). %% https://ui.adsabs.harvard.edu/abs/1986A%26A...157..223D/abstract
    }
    \label{fig:Fig3_Leptonic}
\end{figure*}

\begin{figure*}
    \begin{center}
    \includegraphics[width=16.5cm]{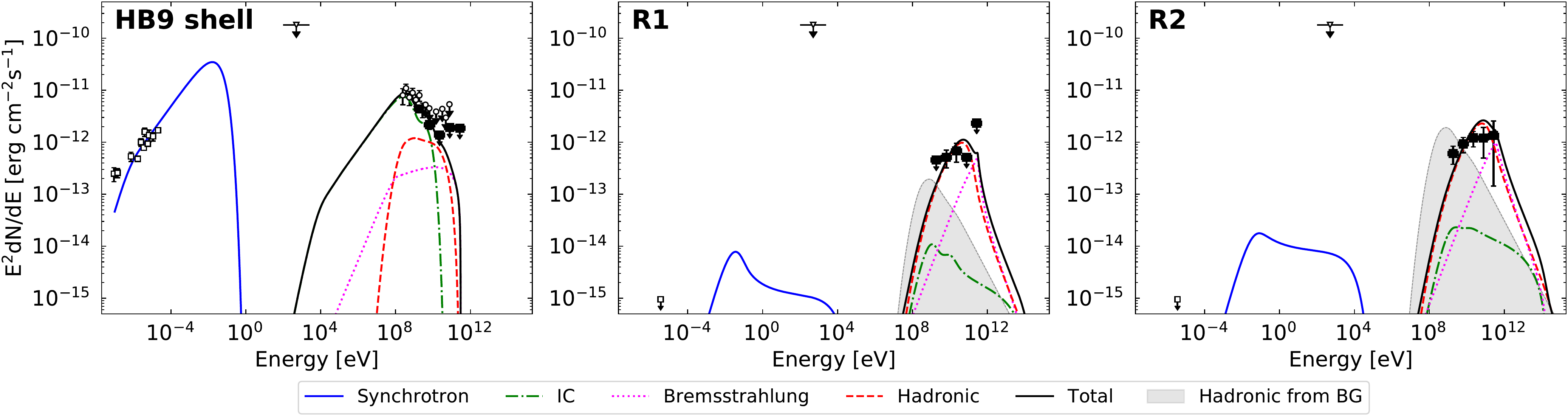}
    \end{center}
    \caption{
    Same as Figure~\ref{fig:Fig3_Leptonic}, but with the hadronic-dominated model.
    }
    \label{fig:Fig4_Hybrid}
\end{figure*}

We fit the spectra of R1 and R2 under two assumptions, the leptonic-dominated and hadronic-dominated models, and simultaneously reproduce the spectrum of the SNR shell with the leptonic emission.
The electron to proton flux ratio, $K_{\rm ep}$, is assumed to be $1$ and $0.02$ for the leptonic-dominated and hadronic-dominated models, respectively, the latter of which is consistent with the ratio in the local CR abundance~(e.g., \citet{2016PhRvL.117i1103A})
\footnote{
% Here $K_{\rm ep}$ is the ratio of CR fluxes in the sources (i.e., in the SNR shell), which does not necessarily correspond to the flux ratio of CR proton and electron measured in the vicinity of the solar system.
Here $K_{\rm ep}$ is the flux ratio of CR proton and electron in the sources (i.e., in the SNR shell), which does not necessarily correspond to the measurements in the solar system.
% In our model, for example, by integrating Equation (\ref{eq:f_oute}) over the whole space, we can obtain the energy spectrum $N_{\rm e,esc}(t,E)$ of the total escaped CR electrons, namely
In our model, for example, by integrating Equation (\ref{eq:f_oute}) over the whole space, we can obtain the energy spectrum $N_{\rm e,esc}(t,E)$ of the total escaped CR electrons, namely
%%%%%%%%%%%%%%%%%%%%%%%%%%%%%%%%%%%%%
\begin{equation}
    N_{\rm e,esc}(t,E)=\int 4\pi r^2 dr f_{\rm e,out}=\frac{K_{\rm ep}N_{\rm esc}(E_c)}{1-Q_{\mathrm{syn}} t_{c} E_{c} / \alpha}\frac{E_c^2}{E^2}.
\end{equation}
%%%%%%%%%%%%%%%%%%%%%%%%%%%%%%%%%%%%%
At the time $t=t_{\rm age}$ (i.e., $E_{\rm esc}(t)=E_{\rm now}$), the value of $N_{\rm e,esc}$ is $N_{\rm e,esc}(t_{\rm age},E_{\rm now})=K_{\rm ep}N_{\rm esc}(E_{\rm now})/\left(1-Q_{\mathrm{syn}}t_{\rm age}E_{\rm now}/\alpha\right)$, which is not identical to the value at the SNR shell, i.e., $N_{\rm e,shell}(t_{\rm age},E_{\rm now})=K_{\rm ep}N_{\rm esc}(E_{\rm now})\neq N_{\rm e,esc}(t_{\rm age},E_{\rm now})$.
% This difference is due to the effect of the radiative cooling in the interstellar space, that is, the flux at the source and the flux of escaped particles do not necessarily match.
This difference is due to the radiative cooling effect in the interstellar space, hence, the flux at the source does not necessarily match with one of the escaped particles.
% Therefore, when comparing the value of $K_{\rm ep}$ to the observed value, to be precise, such a propagation effect should be taken into account.
On the other hand, the propagation effect should be taken into account if comparing the value of $K_{\rm ep}$ with the measurements in the solar system.
However, the factor $1/\left(1-Q_{\mathrm{syn}}t_{\rm age}E_{\rm now}/\alpha\right)$ is almost unity in the parameter range in this paper, thus here we compare $K_{\rm ep}$ directly to the observed value at the earth.
}.
Figure~\ref{fig:Fig3_Leptonic} and~\ref{fig:Fig4_Hybrid} show the results of the leptonic-dominated and hadronic-dominated model, respectively.
The total electron energy and the magnetic field in the SNR shell are in agreement with those estimated by \citet{Araya:2014kra}, while the electron maximum energy at the shell (corresponding to $E_{\rm now}$) is determined to be $300~{\rm GeV}$.
Our obtained value of $E_{\rm now}$ is slightly lower than that derived in the previous study~\citep{Araya:2014kra} because the FIR radiation as a seed photon in the IC process is newly taken into account in this work.
In the leptonic-dominated model, the gamma-ray emissions via a bremsstrahlung process of relativistic electrons dominate in the cloud regions, and thus the delayed gamma-ray spectra for 1--500 GeV have a hard index of $\sim$ 1.3, which contradicts the observed one at the cloud regions (Figure~\ref{fig:Fig3_Leptonic}).
In the hadronic-dominated model (Figure~\ref{fig:Fig4_Hybrid}), the hadronic emission reproduces well the observed spectra even though the assumed parameters in the calculation are typical ones for an SNR and ISM (Table 2).
We note that the gamma-ray flux from the background CRs is lower than the observed data at higher energy than $\sim1$ GeV (see the middle and right panels of Figure~\ref{fig:Fig4_Hybrid}). 
By calculating the energy of the proton corresponding to the peak of photon spectra in each cloud region with the fiducial parameters, we estimate the epoch at which those protons escaped from the SNR shock by using equation (\ref{eq:t_esc}); they are $t_{\rm ref}$ $\lesssim-1.7\times10^{2}$~yr for R1 and $\lesssim-7.5\times10^{2}$~yr for R2 dating back from now.


\subsection{Implications to the diffusion coefficient} \label{section:Modeling3}

\begin{figure*}
    \begin{tabular}{c}
    \begin{minipage}{0.45\hsize}
        \centering
        \includegraphics[width=15cm]{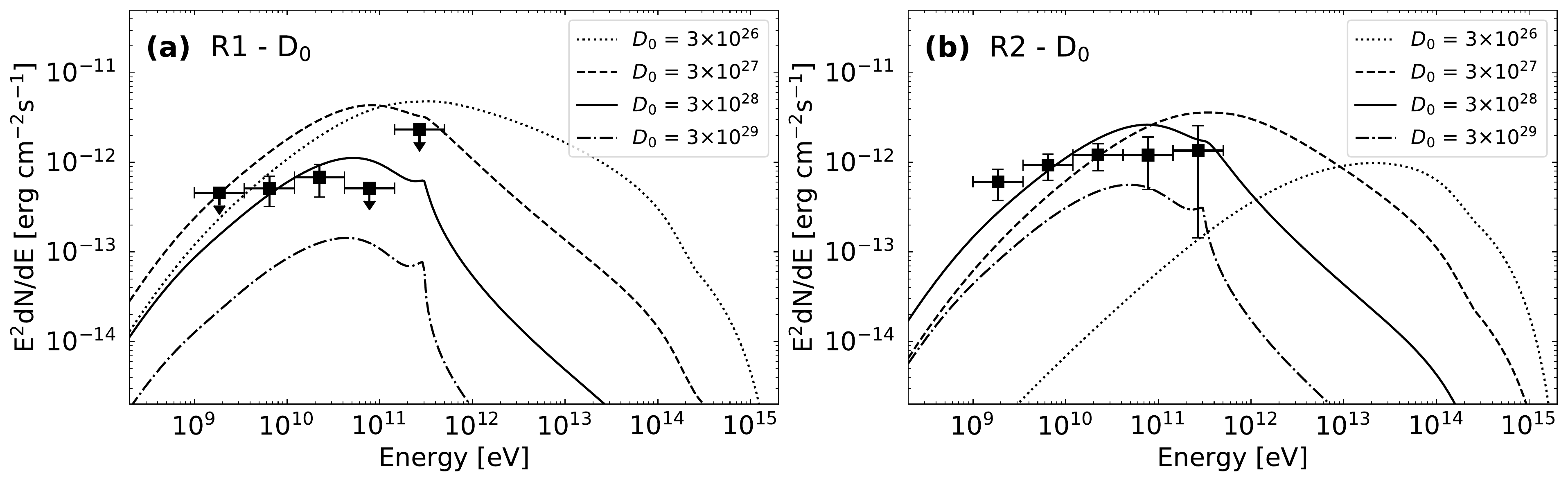}
    \end{minipage}
    \\
    \begin{minipage}{0.45\hsize}
        \centering
        \includegraphics[width=15cm]{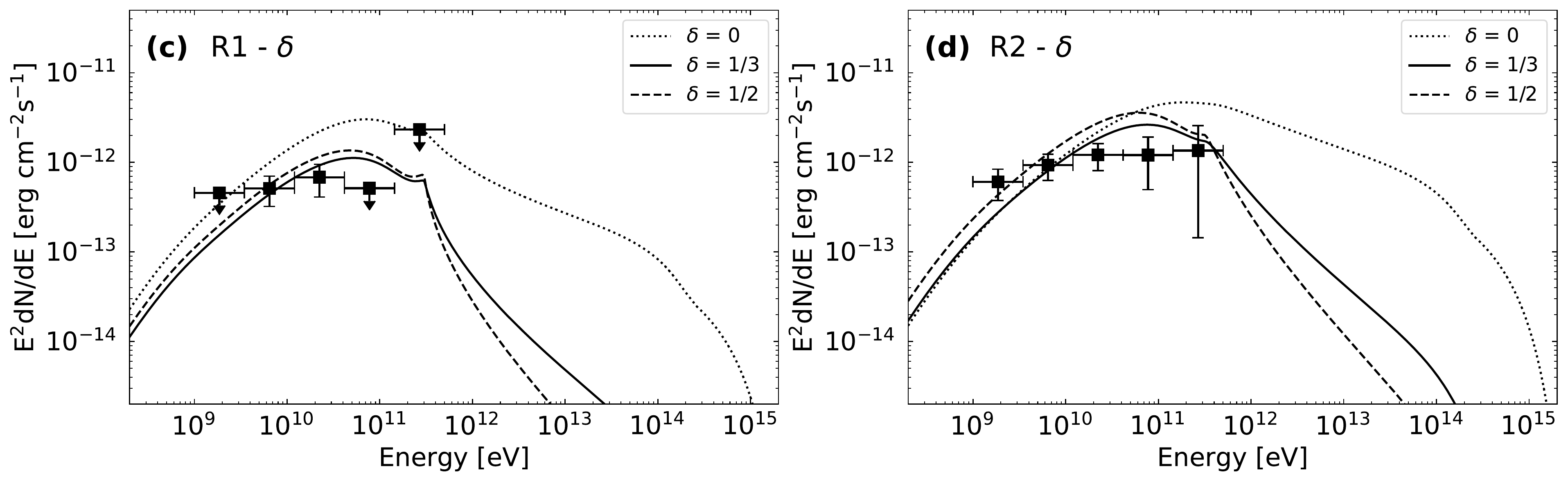}
    \end{minipage}
    \\
    \begin{minipage}{0.45\hsize}
        \centering
        \includegraphics[width=15cm]{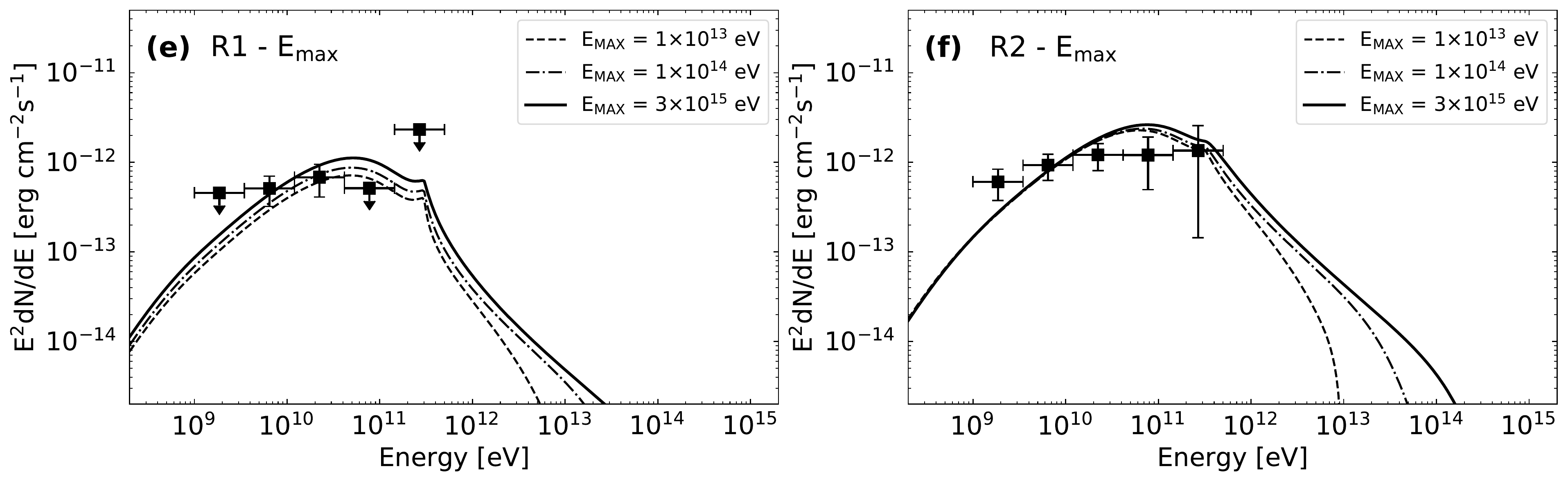}
    \end{minipage}
    \\
    \begin{minipage}{0.45\hsize}
        \centering
        \includegraphics[width=15cm]{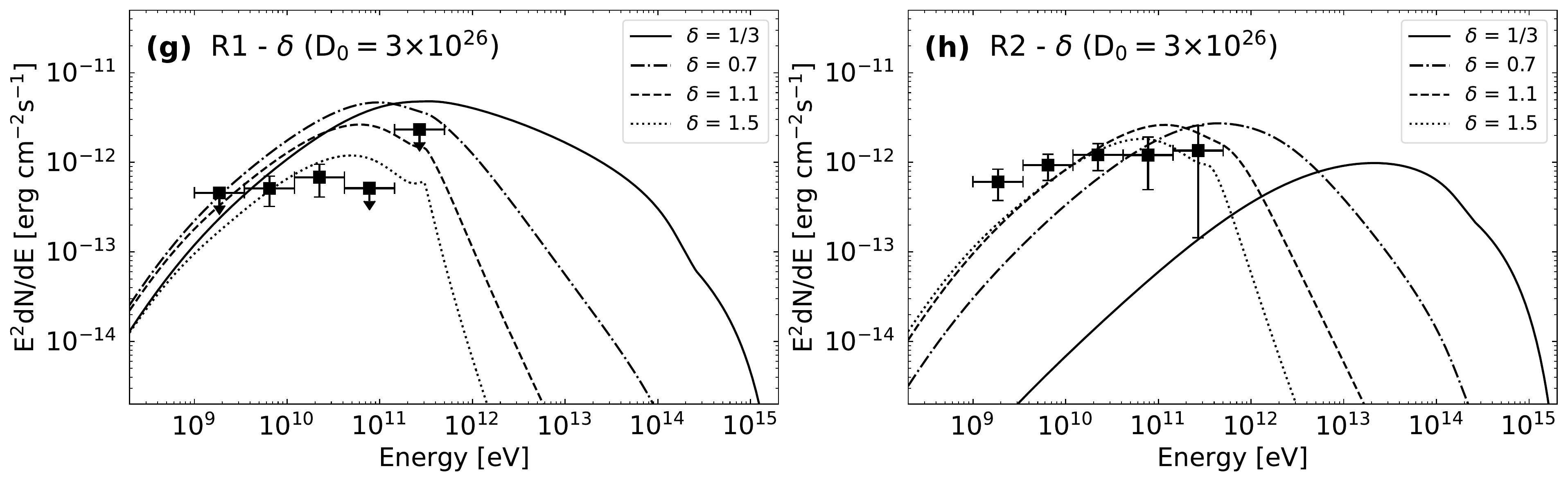}
    \end{minipage}
    \\
    \\
    \end{tabular}
    \caption{
    Input parameter dependency of the model. 
     In panels (a--f), the left (a, c, e, g) and right (b, d, f, h) panels show the results about R1 and R2, respectively.
    {\bf(a, b)}  Dependency on $D_{0}$ (diffusion coefficient). 
    Dotted, dashed, solid, and dot-dashed curves indicate the models with $D_{0}$ of 3 $\times$ 10$^{26}$~cm$^{2}$s$^{-1}$, 3 $\times$ 10$^{27}$~cm$^{2}$s$^{-1}$, 3 $\times$ 10$^{28}$~cm$^{2}$s$^{-1}$, and 3 $\times$ 10$^{29}$~cm$^{2}$s$^{-1}$, respectively.
    {\bf(c, d)}  Dependency on $\delta$ (degree of dependence of $D_0$ on the energy). Dashed, solid, and dot-dashed curves indicate the models with $\delta$ of 0, 1/3, and 1/2, respectively.
    {\bf(e, f)}  Dependency on $E_{\max}$ (maximum energy of the accelerated particles in the Sedov phase).
    Dashed, dot-dashed, and solid curves indicate the models with $E_{\max}$ of 1 $\times$ 10$^{13}$~eV, 1 $\times$ 10$^{14}$~eV, and 3 $\times$ 10$^{15}$~eV, respectively.
    {\bf(g, h)} Dependency on $\delta$ with $D_{0}$ of 3 $\times$ 10$^{26}$~cm$^{2}$s$^{-1}$. Solid, dot-dashed, dashed, and dotted curves indicate the models with $\delta$ of 1/3, 0.7, 1.1, and 1.5, respectively.
    }
    \label{fig:Fig4_ModelAll}
\end{figure*}

We investigate in the following procedure how the hadronic-dominated model curve varies depending on the input parameters and summarize the results in Figure~\ref{fig:Fig4_ModelAll}.
First, we evaluate the dependence of the model curve on $D_0$, which is the value of the diffusion coefficient at $E=10~{\rm GeV}$ (Figures \ref{fig:Fig4_ModelAll}a and \ref{fig:Fig4_ModelAll}b).
The \textit{Fermi}-LAT spectra obtained in this  work are found to be well reproduced with the Galactic mean of $D_0 = 3 \times 10^{28}~{\rm cm^2~s^{-1}}$.
Orders of magnitude smaller $D_0$, in particular $D_0=3\times 10^{26}~{\rm cm^2~s^{-1}}$,  however, clearly fail to reproduce the observed spectra.
In the previous studies for other SNRs~(e.g., \citet{Fujita:2009ak, Acciari:2020toc}), the estimated values of $D_{0}$ were $\sim 10$ times smaller than the Galactic mean, which were explained in conjunction with self-confinement caused by the generation of turbulent plasma waves~\citep{Wentzel:1974cp, Fujita:2011wq, DAngelo:2017rou}.
Our diffusion coefficient value, which is close to the Galactic mean, indicates that the excitation of such turbulent plasma waves at the distances to R1 and R2 is inefficient or that the wave damping has a significant effect.
Second, we find that the model with either $\delta = 1/3$  or $1/2$ is preferred over that with $\delta = 0$ (Figures~\ref{fig:Fig4_ModelAll}c and \ref{fig:Fig4_ModelAll}d).
Third, we also find that $E_{\max}$ above 10~TeV still explains the data points{ (Figures~\ref{fig:Fig4_ModelAll}e and \ref{fig:Fig4_ModelAll}f).}
Given that there is a trend for a larger difference between the model curves in the higher energy band, future observations in the TeV band  will provide results more sensitive to determine the $\delta$ and $E_{\max}$ parameters.

In order to verify that the Galactic mean value of the diffusion coefficient ($D_{0}$) is appropriate to explain the gamma-ray spectra, we also try to model the gamma-ray spectra, fixing $D_{0}$ at two orders of magnitude smaller $D_0=3\times 10^{26}~{\rm cm^2~s^{-1}}$.
As can be seen in Figure \ref{fig:Fig4_ModelAll}(a, b), for smaller $D_0$, the spectrum shifts to the higher energy side.
Also, as shown in Figure \ref{fig:Fig4_ModelAll}(e, f), the spectrum below $\mathcal{O} \left(10^{11}\right)$ eV does not depend much on $E_{\rm max}$.
Therefore, the deviation of the spectrum in the lower energy band due to the small $D_0$ should be countered by $\delta$ dependence.
Once we assume $\delta > 1.0$ while other parameters are kept the same as in Table~\ref{tab:modelpara}, the model curves are roughly consistent with the observed gamma-ray spectra (Figure~\ref{fig:Fig4_ModelAll}(g, h)).
However, $\delta > 1.0$ is inconsistent with the CR propagation model~(e.g., \citet{Blasi:2011fm}).
Thus, the small diffusion coefficient (i.e., $D_0\sim 10^{26}~{\rm cm^2~s^{-1}}$) is unlikely.

\begin{figure*}
    \begin{center}
    \includegraphics[width=16.5cm]{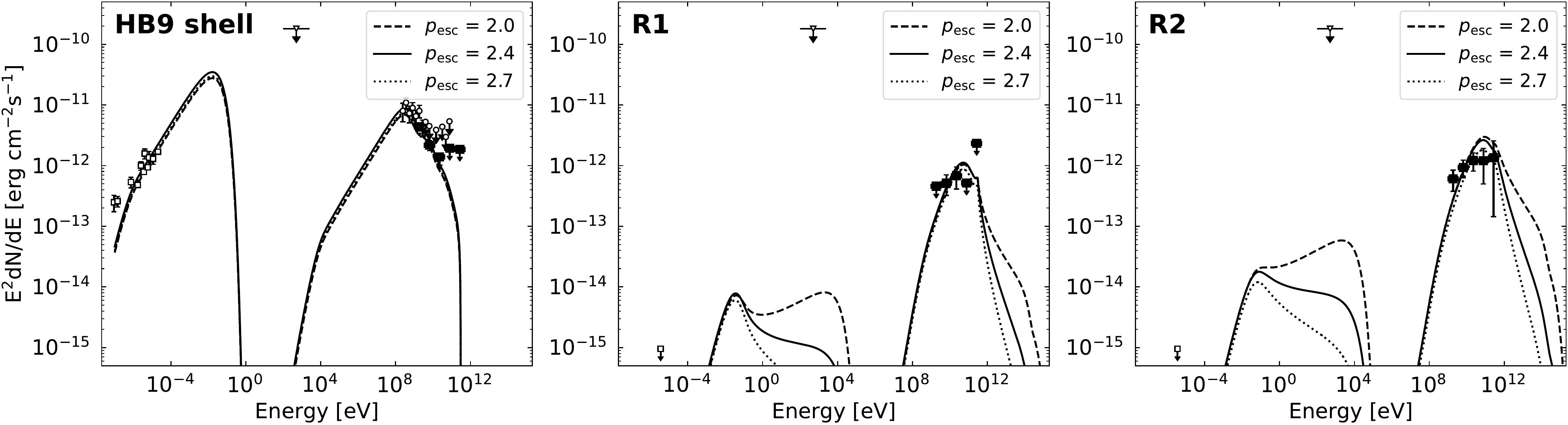}
    \end{center}
    \caption{
    Energy spectra with the hadronic-dominated model under different assumptions on $p_{\rm esc}$.
    The dashed, solid, and dotted lines represent the model emissions with $p_{\rm esc}$ of 2.0, 2.4, and 2.7, respectively.
    The model with $p_{\rm esc} = 2.4$ is the same as in Figure~\ref{fig:Fig4_Hybrid}, while for the models with $p_{\rm esc} = 2.0$ and $2.7$, the acceleration efficiency is given to be $\eta = 0.3$ and $0.05$, respectively.
    %The dashed, solid, and dotted lines represent the total emissions of synchrotron, electron bremsstrahlung, IC, and neutral pion decay.
    }
    \label{fig:Fig6_best_pescdep}
\end{figure*}

\begin{figure*}
    \begin{tabular}{c}
    \begin{minipage}{0.45\hsize}
        \centering
        \includegraphics[width=15cm]{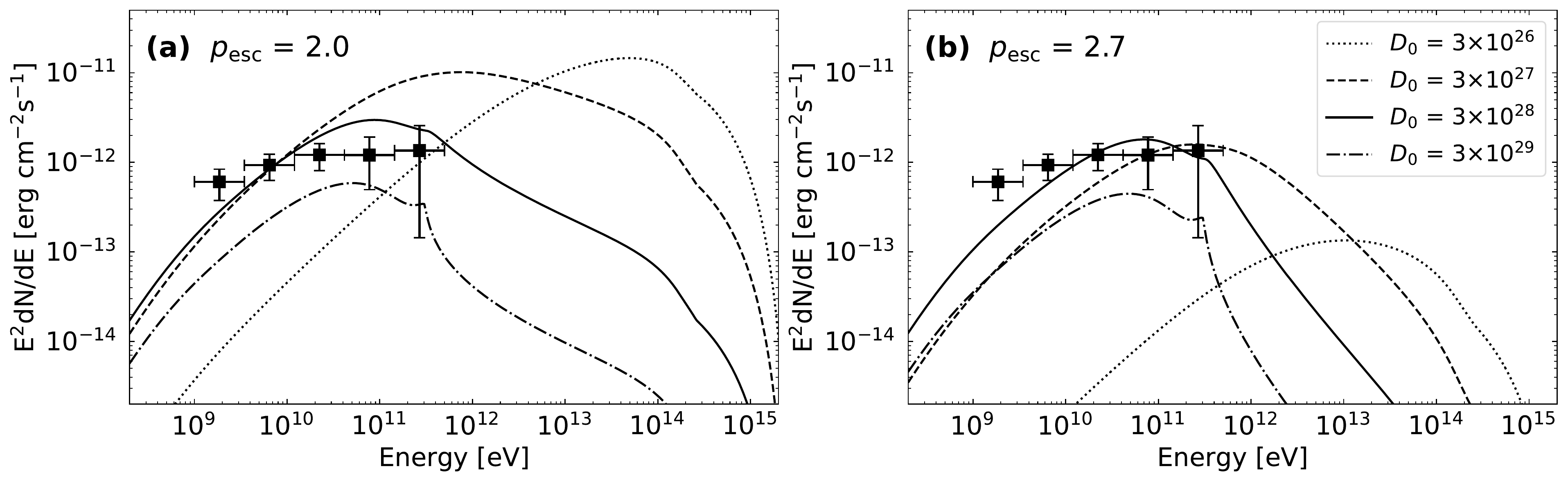}
    \end{minipage}
    \\
    \begin{minipage}{0.45\hsize}
        \centering
        \includegraphics[width=15cm]{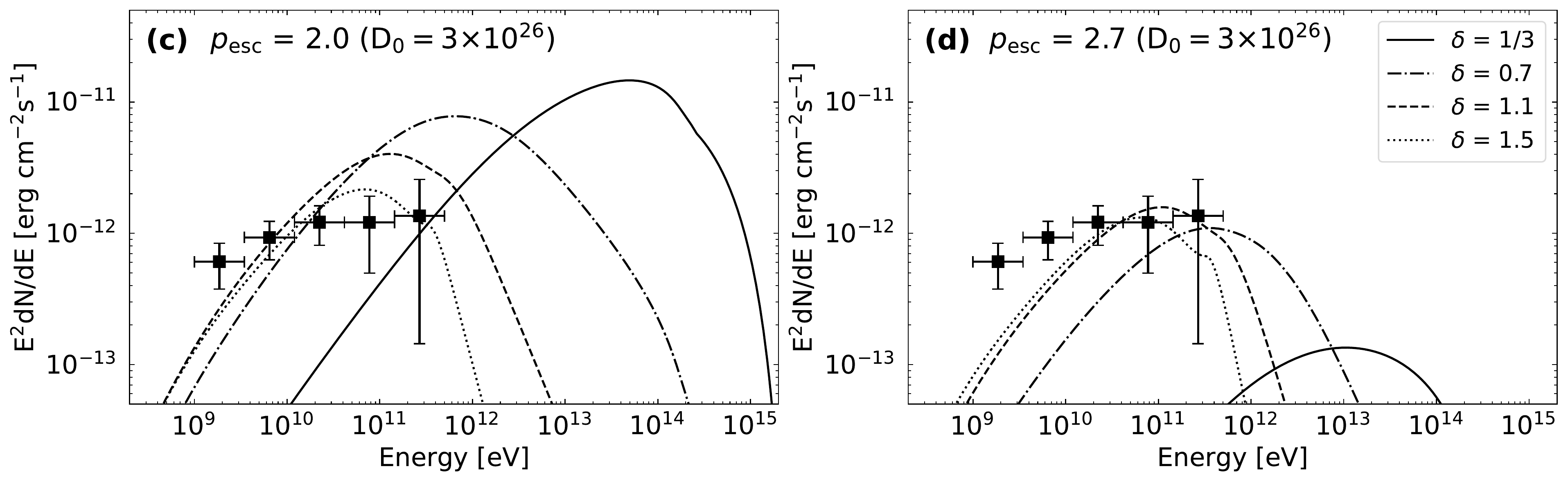}
    \end{minipage}
    \\
    \\
    \end{tabular}
    \caption{
    Input parameter dependency of the model in the R2 region when $p_{\rm esc} = 2.0$ (left) and 2.7 (right) are assumed.
    {\bf(a, b)}  Dependency on $D_{0}$ (diffusion coefficient). 
    %Dotted, dashed, solid, and dot-dashed curves indicate the models with $D_{0}$ of 3 $\times$ 10$^{26}$~cm$^{2}$s$^{-1}$, 3 $\times$ 10$^{27}$~cm$^{2}$s$^{-1}$, 3 $\times$ 10$^{28}$~cm$^{2}$s$^{-1}$, and 3 $\times$ 10$^{29}$~cm$^{2}$s$^{-1}$, respectively.
    {\bf(c, d)} Dependency on $\delta$ fixing at $D_{0}$ = 3 $\times$ 10$^{26}$~cm$^{2}$s$^{-1}$. 
    % Solid, dot-dashed, dashed, and dotted curves indicate the models with $\delta$ of 1/3, 0.7, 1.1, and 1.5, respectively.
    }
    \label{fig:Fig7}
\end{figure*}

We have assumed $p_{\rm esc}=2.4$ in the modeling so far, but the discussion on the diffusion coefficient may be affected by this parameter as well as $\delta$.
Here, we attempt to fit the observed spectra by considering two cases where a flatter ($p_{\rm esc}=2.0$) or steeper ($p_{\rm esc}=2.7$) index is assumed.
Note that these assumptions are not favored by the propagation model as mentioned in Section~\ref{section:Modeling2}.
The results are shown in Figure~\ref{fig:Fig6_best_pescdep} and \ref{fig:Fig7}.
Although the model parameters except for $p_{\rm esc}$ and $\eta$ are fixed at the values in Table~\ref{tab:modelpara}, the observed spectra can be reproduced by giving a feasible value of $\eta$ = 0.3 (0.05) for $p_{\rm esc}$ = 2.0 (2.7) (Figure~\ref{fig:Fig6_best_pescdep}).
Figure~\ref{fig:Fig7} shows the dependency on $D_0$ and $\delta$ for R2 as in Figure~\ref{fig:Fig4_ModelAll} (b) and (h).
To explain the observed spectrum with the diffusion coefficient two orders of magnitude smaller $D_0=3\times 10^{26}~{\rm cm^2~s^{-1}}$ than the Galactic mean, an unrealistic assumption that $\delta > 1.0$ is required for both cases of $p_{\rm esc}=2.0$ and $2.7$.
This result is consistent with the case where $p_{\rm esc} = 2.4$, i.e., the diffusion coefficient of the Galactic mean value is preferred regardless of the assumption on $p_{\rm esc}$.

\section{Conclusions} \label{section:Summary}

We analyzed the GeV gamma-ray emissions in the vicinity of SNR HB9 with the \textit{Fermi}-LAT  data spanning for 12 years, aiming to quantify the  evolution of  DSA.
We detected significant gamma-ray emission spatially coinciding with  two MCs in the vicinity of the SNR.
We found that the gamma-ray spectra above 1~GeV at the cloud regions could be characterized by a simple power-law function with indices of $1.84 \pm 0.18$ and $1.84 \pm 0.14$, which are flatter than that at the SNR shell of $2.55 \pm 0.10$.
By modeling the diffusion of the CRs that escaped from SNR HB9, we found that the gamma-ray emission could be explained with the scenario that both CR protons and electrons accelerated by the SNR illuminate MCs rather than the leptonic scenario (see Figures \ref{fig:Fig3_Leptonic} and \ref{fig:Fig4_Hybrid}).
This result implies that the maximum energy with DSA in younger SNRs is likely to be higher than  that in older ones.
We also found that the Galactic mean value of the diffusion coefficient ($D_{0}$) is appropriate to explain the observed gamma-ray spectra, indicating that self-confinement by turbulent plasma waves is not effective in the vicinity of SNR HB9.
Future observations in the TeV band will provide results more sensitive to determine the maximum energy of the SNR in the past.

%%%%%%%%%%%%%%%%%%%%%%%%%%%%%%%%%%%%%%%%%%%%%%%%%%%%%%

\begin{ack}
% acknowledgments
We would like to thank Toshihiro Fujii, Yutaka Fujita, Hidetoshi Kubo, Takaaki Tanaka, and Kenta Terauchi for helpful discussion and comments.
We thank the participants and the organizers of the workshops with the identification number YITP-W-20-01 for their generous support and helpful comments.
This work is supported by a Grant-in-Aid for JSPS Fellows Grant No. 20J21480 (TO).
\end{ack}

\appendix

\section{Derivation of the spectrum of CR electrons} \label{Appendix:derivation}

The transport equation for CR electrons is
%%%%%%%%%%%%%%%%%%%%%%%%%%%%%%%%%%%%%
%\begin{equation}\label{eq:transport_e_app}
\begin{eqnarray}\label{eq:transport_e_app}
    \frac{\partial f_{e,\rm out}}{\partial t}(t, r, E)
    && - D_{\mathrm{ISM}}(E) \Delta f_{e,\rm out}(t, r, E) \nonumber \\
    && + \frac{\partial}{\partial E}\left(P(E)f_{e,\rm out}(t, r, E)\right)
    %=q_{e,s}(t, r, E).
    =q_{e,s}(t, r, E).
%\end{equation}
\end{eqnarray}
%%%%%%%%%%%%%%%%%%%%%%%%%%%%%%%%%%%%%
According to \citet{1995PhRvD..52.3265A}, the Green function of equation (\ref{eq:transport_e_app}) is
%%%%%%%%%%%%%%%%%%%%%%%%%%%%%%%%%%%%%
\begin{eqnarray}\label{eq:Green_cooling}
&& G(t, \mathbf{r}, E;t_0,\mathbf{r}_0;E_0) \nonumber \\
 && = \frac{P\left(E_{t-t_0}\right)}{\pi^{3 / 2} P(E) r_{\mathrm{dif}}^{3}} \exp \left(-\frac{\left|\mathbf{r}-\mathbf{r}_0\right|^{2}}{r_{\mathrm{dif}}^{2}}\right) \delta\left(E_{t-t_0}-E_0\right),  
\end{eqnarray}
%%%%%%%%%%%%%%%%%%%%%%%%%%%%%%%%%%%%%
where $E_{t-t_0}$ is defined by the following implicit relation:
%%%%%%%%%%%%%%%%%%%%%%%%%%%%%%%%%%%%%
\begin{equation}\label{eq:Ett0}
   t-t_0=\int_{E}^{E_{t-t_0}} \frac{d E'}{P(E')},
\end{equation}
%%%%%%%%%%%%%%%%%%%%%%%%%%%%%%%%%%%%%
%%%%%%%%%%%%%%%%%%%%%%%%%%%%%%%%%%%%%
\begin{equation}
    r_{\mathrm{dif}}=\sqrt{4\Delta u},
\end{equation}
%%%%%%%%%%%%%%%%%%%%%%%%%%%%%%%%%%%%%
and
%%%%%%%%%%%%%%%%%%%%%%%%%%%%%%%%%%%%%
\begin{equation}\label{eq:deltau}
    \Delta u=\int_{E}^{E_{t-t_0}} \frac{D_{\rm ISM}(E')}{P(E')} d E'.
\end{equation}
%%%%%%%%%%%%%%%%%%%%%%%%%%%%%%%%%%%%%
From linearity, the solution to equation (\ref{eq:transport_e_app}) can be calculated as follows:
%%%%%%%%%%%%%%%%%%%%%%%%%%%%%%%%%%%%%
\begin{equation} 
\label{eq:linearity}
    f_{\rm{e, out}}(\mathrm{r}, t, E)=
    \int d E_0 \int d^{3} \mathrm{r}_{0} \int d t_{0}
    G(t, \mathbf{r}, E;t_0,\mathbf{r}_0;E_0)
    q_{e,s}\left(\mathrm{r}_{\mathbf{0}}, t_{0}, E_0\right).    
\end{equation}
%%%%%%%%%%%%%%%%%%%%%%%%%%%%%%%%%%%%%

In this paper, only synchrotron radiation is considered as the most important process in radiative cooling, so the cooling function $P(E)$ can be written as follows:
%%%%%%%%%%%%%%%%%%%%%%%%%%%%%%%%%%%%%
\begin{equation}\label{eq:syn_cooling}
    P(E)=\frac{4}{3}\sigma_T c \left(\frac{E}{m_e c^2}\right)^2 \frac{B_{\rm ISM}^2}{8\pi}\equiv Q_{\rm syn}E^2,
\end{equation}
%%%%%%%%%%%%%%%%%%%%%%%%%%%%%%%%%%%%%
where $\sigma_T$ is the Thomson scattering cross section, $m_e$ is the electron mass, and $B_{\rm ISM}$ is a value of the magnetic field in the interstellar space.
Evaluating equation (\ref{eq:Ett0}) with using equation (\ref{eq:syn_cooling}), we obtain 
%%%%%%%%%%%%%%%%%%%%%%%%%%%%%%%%%%%%%
\begin{equation}\label{eq:coef_syn1}
    % E_{t-t_0}=\frac{E}{1-P(E)(t-t_0)/E}.
    E_{t-t_0}=\frac{E}{1-Q_{\rm syn}E(t-t_0)}.
\end{equation}
%%%%%%%%%%%%%%%%%%%%%%%%%%%%%%%%%%%%%
For equation (\ref{eq:deltau}), we can calculate using equation (\ref{eq:DISM}) and (\ref{eq:syn_cooling}) as follows:
%%%%%%%%%%%%%%%%%%%%%%%%%%%%%%%%%%%%%
\begin{equation}\label{eq:coef_syn2}
    \Delta u=\frac{r_{\rm dif}^2}{4}
    =
    \frac{D_{\rm ISM}(E)}{\left(1-\delta\right)Q_{\rm syn}E}
    \left[
    1-\left(\frac{E}{E_{t-t_0}}\right)^{1-\delta}
    \right].
\end{equation}
%%%%%%%%%%%%%%%%%%%%%%%%%%%%%%%%%%%%%

By using equations (\ref{eq:Ecut}), (\ref{eq:qs}), (\ref{eq:qse}), (\ref{eq:Green_cooling}), (\ref{eq:coef_syn1}), and (\ref{eq:coef_syn2}) and performing the integration of equation (\ref{eq:linearity}), the solution to equation (\ref{eq:transport_e_app}) can be obtained as follows:
%%%%%%%%%%%%%%%%%%%%%%%%%%%%%%%%%%%%%
%\begin{equation}
\begin{eqnarray}
%f_{\rm {e,out}}(t, r, E)&=&\frac{K_{\rm ep}N_{\mathrm{esc}}\left(E_{c}\right)}{4 \pi^{3 / 2} r R_{\mathrm{c}} R_{\rm d, e}} \frac{E_{c}^{2}}{E^{2}} \frac{1}{1-Q_{\rm syn}t_{c}E_c / \alpha} \left[\exp \left(-\frac{\left(r-R_{\mathrm{c}}\right)^{2}}{R_{\rm d, e}^{2}}\right)-\exp \left(-\frac{\left(r+R_{\mathrm{c}}\right)^{2}}{R_{\rm d, e}^{2}}\right)\right],
f_{\rm {e,out}}(t, r, E) = \frac{K_{\rm ep}N_{\mathrm{esc}}\left(E_{c}\right)}{4 \pi^{3 / 2} r R_{\mathrm{c}} R_{\rm d, e}} \frac{E_{c}^{2}}{E^{2}} \frac{1}{1-Q_{\rm syn}t_{c}E_c / \alpha}  \nonumber \\ ~\times \left[\exp \left(-\frac{\left(r-R_{\mathrm{c}}\right)^{2}}{R_{\rm d, e}^{2}}\right)-\exp \left(-\frac{\left(r+R_{\mathrm{c}}\right)^{2}}{R_{\rm d, e}^{2}}\right)\right],
%\end{equation}
\end{eqnarray}
%%%%%%%%%%%%%%%%%%%%%%%%%%%%%%%%%%%%%
where $t_c=t_c(t,E)$ is the time when the electron whose energy is $E$ at the current time $t$ is injected and determined as a solution to an algebraic equation $t_c=t_{\rm esc}(E_{t-t_c})$, or equivalently the below equation,
%%%%%%%%%%%%%%%%%%%%%%%%%%%%%%%%%%%%%
\begin{equation} \label{eq:tceqs}
    % $E_c=E_{t-t_c}=E_{\rm esc}(t_c)$, 
    \frac{E}{1-Q_{\rm syn}E\left(t-t_c\right)}=E_{\rm esc}(t_c),
\end{equation}
%%%%%%%%%%%%%%%%%%%%%%%%%%%%%%%%%%%%%
$E_c(t,E)=E_{\rm esc}(t_c)$, $R_c(t,E) \equiv R_{\mathrm{esc}}\left(E_{c}\right)$, and $R_{\rm d, e}(t,E)$ is the diffusion length of CR electrons
%\widel{$R_{\rm d, e}\equiv r_{\mathrm{dif}}\left(E_{c}\right)$.}
%%%%%%%%%%%%%%%%%%%%%%%%%%%%%%%%%%%%%
\begin{equation}
    R_{\rm d, e}=\sqrt{\frac{4D_{\rm ISM}(E)}{\left(1-\delta\right)Q_{\rm syn}E}
    \left[
    1-\left(\frac{E}{E_c}\right)^{1-\delta}
    \right]}.
\end{equation}
%%%%%%%%%%%%%%%%%%%%%%%%%%%%%%%%%%%%%
Here, we have assumed that there is only one solution to equation (\ref{eq:tceqs}), however this is not generally true.
The condition for there to be only one solution can be written as follows:
%%%%%%%%%%%%%%%%%%%%%%%%%%%%%%%%%%%%%
\begin{equation}
    t_{\rm Sedov} \geq t_{c, \rm{ crit }}
    \equiv t_{\rm {Sedov }}\left[\frac{Q_{\rm syn} t_{\rm {Sedov }} E_{\rm max}}{\alpha}\right]^{1 /(\alpha-1)}
\end{equation}
%%%%%%%%%%%%%%%%%%%%%%%%%%%%%%%%%%%%%
For the range of parameters we use in this paper, this condition is satisfied, and then the assumption of only one solution is consequently justified.

\end{document}